\documentstyle [12pt,epsfig]{article} 
\makeatletter
\@addtoreset{equation}{section}
\makeatother

\newcommand{\BS}{\bigskip}
\newcommand{\SECTION}[1]{\BS{\large\section{\bf #1}}}

\begin{document}
\begin{titlepage}
\begin{center}
\vspace*{2cm}
{\large \bf
  The Description of Neutrino and Muon Oscillations by
 Interfering Amplitudes of Classical Space-Time Paths  }
\vspace*{1.5cm}
\end{center}
\begin{center}
{\bf J.H.Field }
\end{center}
\begin{center}
{ 
D\'{e}partement de Physique Nucl\'{e}aire et Corpusculaire
 Universit\'{e} de Gen\`{e}ve . 24, quai Ernest-Ansermet
 CH-1211 Gen\`{e}ve 4.
}
\end{center}
\vspace*{2cm}
\begin{abstract}
 Particle oscillations following neutrino production processes are 
  analysed within Feynman's path amplitude formulation of quantum
  mechanics. Consideration of the temporal sequence of production
  and detection events reveals an important contribution to the
  oscillation phase from the space-time propagator of the decaying
  source particle. Formulae are derived for spatial neutrino 
   oscillations following pion decay (at rest and in flight), muon
  decay and nuclear $\beta$-decay at rest, as well as for similar muon 
  oscillations after pion decay at rest and in flight. In all cases
   studied, the neutrino oscillation phases found differ from
  the conventionally used one. A concise critical
  review is made of previous treatments of the quantum mechanics
  of neutrino and muon oscillations.                     
\end{abstract}
\vspace*{1cm}
\newline
{\it Keywords ;} Quantum Mechanics,
Neutrino Oscillations.
\end{titlepage}
  
\SECTION{\bf{Introduction}}
The quantum mechanical description of neutrino oscillations~\cite{Pont1,Pont2}
has been the subject of much discussion and debate in the recent literature. The
`standard' oscillation formula~\cite{BiPet}, yielding an oscillation phase\footnote{
 The interference term is proportional to $\cos \phi_{12}$ or
 $\sin^2 \frac{\phi_{12}}{2}$.}, at distance $L$ from the neutrino source, between
 neutrinos, of mass $m_1$ and $m_2$ and momentum $P$, of: \footnote{Units with $h/2\pi=c=1$
 are used throughout.}
\begin{equation}
 \phi_{12} =\frac{(m_1^2-m_2^2)L}{2P},
\end{equation}
is derived on the assumption of equal momentum at production of the two
neutrino mass eigenstates. Other authors have proposed, instead, equal energies
~\cite{GroLip} or velocities~\cite{LDR} at production, confirming, in both cases,
 the result of the standard formula. The latter reference claims, however, that
 the standard expression for $\phi_{12}$ should be multiplied by a factor of two in the 
 case of the equal energy or equal momentum hypotheses. However, the equal momentum,
 energy or velocity assumptions are all incompatible with energy-momentum conservation
 in the neutrino production process~\cite{Win}.
 In two recent calculations~\cite{Moh1,SWS} a covariant formalism was used in which exact
 energy-momentum conservation was imposed. Using the invariant Feynman
 propagator~\cite{Feyn1} to describe the space-time evolution of the neutrino mass
 eigenstates, the author of Ref.~\cite{Moh1} also found an oscillation phase a factor
 of two larger than given by the the standard formula. However, on convoluting
 the transition amplitude with a Gaussian wavefunction descibing the spatial-temporal
 position of the source pion, the standard result was recovered. In Ref.~\cite{SWS},
 a completely different formula was obtained for the neutrino oscillation phase following
 pion decay and it was predicted that only correlated oscillations of neutrinos and
 the recoiling decay muons could be observed. However, the author of Ref.~\cite{Moh1}
 as well as others~\cite{BLSG,DMOS} claimed that muon oscillations would either be 
 completely suppressed, or essentially impossible to observe.
 
 \par The present paper calculates the probabilities of oscillation of neutrinos and muons
 produced by pions decaying both at rest and in flight, as well as the probabilities of 
neutrino oscillation following muon decay or
 $\beta-$decay of a nucleus at rest. The calculations, which are fully covariant, are based on
 Feynman's reformulation of quantum mechanics~\cite{Feyn2} in terms of interfering amplitudes
 associated with classical space-time particle trajectories. The essential interpretational
 formula of this approach\footnote{Postulate 1 and Eqn.(7) of Ref.~\cite{Feyn2}.}, though
 motivated by the seminal paper of Dirac on the Lagrangian formulation of quantum mechanics
 ~\cite{Dirac}, and much developed later in the work of Feynman and other authors~\cite{Feyn3},
 was actually already given by Heisenberg in 1930
 \footnote{Heisenberg remarked that the fundamental formula (1.2) must be 
   distinguished from that where the summation over intermediate states is made at the level
   of probabilities, rather than amplitudes, and that the distinction between
  the two formulae is `the centre of the whole quantum theory'.}~\cite{Heis}. The application of the 
 path amplitude formalism to neutrino or muon oscillations is particularly staightforward,
 since, in the covariant formulation of quantum mechanics, energy and momentum are 
  exactly conserved at all vertices and due to the macroscopic propagation distances of
  the observed neutrinos and muons all these particles follow essentially classical
  trajectories ({\it i.e.} corresponding to the minima of the classical action) which
  are rectilinear paths with constant velocities. The essential formula of 
  Feynman's version of quantum mechanics, to be employed in the calculations 
  presented below, is~\cite{Feyn2,Heis}:
 \begin{equation}
 P_{fi} = \left|\sum_{k_1} \sum_{k_2}...\sum_{k_n}\langle f| k_1 \rangle
\langle k_1| k_2 \rangle...\langle k_n| i \rangle \right|^2  
 \end{equation}
where $P_{fi}$ is the probability to observe a final state $f$, given an initial
 state $i$, and $k_j,~j=1,n$ are (unobserved) intermediate quantum states.
 In the applications to be described in this paper, Eqn(1.2) specialises to
  \footnote{In Eqns(1.3) and (1.4) an additional summation over
   unobserved states, with different physical masses of the decay muon, is omitted for
   simplicity. See Eqns.(2.1) and (2.35) below.}:
 \begin{equation}
 P_{e^- \pi^+} = \left|\sum_{k=1,2}\langle e^-| \nu_k \rangle
\langle \nu_k,x_D | \nu_k, x_k \rangle \langle \nu_k | \pi^+ \rangle
 \langle \pi^+,x_k |  \pi^+, x_0 \rangle \langle \pi^+,x_0 | S_{\pi}\rangle \right|^2  
 \end{equation}
for the case of neutrino oscillations and 
 \begin{equation}
 P_{e^+ \pi^+} = \left|\sum_{k=1,2}\langle e^+| \mu_k^+ \rangle
\langle \mu_k^+,x_D | \mu_k^+, x_k \rangle \langle \mu_k^+ | \pi^+ \rangle
 \langle \pi^+,x_k |  \pi^+, x_0 \rangle \langle \pi^+,x_0 | S_{\pi}\rangle \right|^2  
 \end{equation} 
 for the case of muon oscillations.
  $P_{e^- \pi^+}$ is the probability to observe the charged current neutrino
  interaction: $\nu_e n \rightarrow e^- p$ following the decay:
 $\pi^+ \rightarrow \mu^+ \nu_{\mu}$, while $P_{e^+ \pi^+}$ is the probability
 to observe the decay $\mu^+ \rightarrow e^+ \nu_e \overline{\nu}_{\mu}$, after
 the same decay process. 
 In Eqns.(1.3),(1.4) 
 $| \nu_k \rangle, k=1,2$ are neutrino mass eigenstates
 while $| \mu_k^+ \rangle, k=1,2$ are the corresponding recoil
 muon states from pion decay.  $\langle \nu_k | \pi^+ \rangle$,
 $\langle \mu_k^+ | \pi^+ \rangle$  and $\langle e^+| \mu_k^+ \rangle$ 
 denote invariant decay amplitudes, $\langle e^-| \nu_k \rangle$ is the
 invariant amplitude of the charged current neutrino interaction, 
  $\langle p ,x_2 |  p, x_1 \rangle$ is the invariant space-time propagator
  of particle $p = \nu, \mu, \pi $ between the space-time points $x_1$ and $x_2$
 and $\langle \pi^+,x_0 | S_{\pi}\rangle$ is an invariant amplitude describing
 the production of the $\pi^+$ by the source $S_{\pi}$ and its space-time propagation
  to the space-time point $x_0$.
\par The difference of the approach used in the present paper to previous
 calculations presented in the literature can be seen immediately on
 inspection of Eqns(1.3) and (1.4). The initial state\footnote{ The pion production
 and propagation amplitude $\langle \pi^+,x_0 | S_{\pi}\rangle$ contributes 
 only a multiplicative constant to the transition probabilites. The initial state
 can then just as well be defined as `pion at $x_0$', rather than $| S_{\pi}\rangle$.
  This is done in the calculations presented in Section 2 below.}
 is a pion at space time-point
 $x_0$, the final state an $e^-$ or $e^+$ produced at space-time point $x_D$. These
 are unique points, for any given event and do not depend in any way on the
 masses of the unobserved neutrino eigenstates propagating from $x_k$ to
 $x_D$ in Eqn(1.3). On the other hand the (unobserved) space-time points
 $x_k$ at which the neutrinos and muons are produced {\it do} depend on $k$.
 Indeed, because of the different velocities of the propagating neutrino
 eigenstates, only in this case can both neutrinos and muons ( representing
  {\it alternative} classical histories of the decaying pion) both arrive
  simultanously at the unique point $x_D$ where the neutrino interaction 
  occurs (Eqn(1.3)) or the muon decays (Eqn(1.4)).
 \par The crucial point in the above discussion is that the decaying pion,
  {\it via} the different path amplitudes in Eqns(1.3) and (1.4), {\it
  interferes with itself}. To modify very slightly Dirac's famous
  statement\footnote{`Each photon then interferes only with itself. 
  Interference between two different photons never occurs'~\cite{Dirac1}.}:
 `Each pion then interferes only with itself. 
  Interference between two different pions never occurs'. As will be seen in
  the discussion in Section 5 below, essentially all recent treatments
  of neutrino oscillations in the literature have attempted to 
  contradict the second sentence in Dirac's statement!
  \par Because of the different possible decay times of the pion
   in the two interfering path amplitudes, the pion propagators 
   $\langle \pi^+,x_k |  \pi^+, x_0 \rangle$ in Eqns(1.3) and (1.4) above
    give important contributions to the interference phase. To the 
    author's  best knowledge, this effect has not been taken into 
    account in any previously published calculation of neutrino
    oscillations.  
\par The results found for the oscillation phase are, for pion decays at rest:
 \begin{equation}
\phi_{12}^{\nu,\pi}  = \phi_{12}^{\mu,\pi}  =  \frac{2 m_{\pi} m_{\mu}^2 \Delta m^2 L}
 {(m_{\pi}^2-m_{\mu}^2)^2}   
\end{equation}
and for pion decays in flight:
\begin{eqnarray}
\phi_{12}^{\nu,\pi}& = & \frac{ m_{\mu}^2 \Delta m^2 L}
{(m_{\pi}^2-m_{\mu}^2) E_{\nu} \cos \theta_{\nu}}   \\
 \phi_{12}^{\mu,\pi} & = & \frac{2 m_{\mu}^2 \Delta m^2 ( m_{\mu}^2 E_{\pi}-
m_{\pi}^2 E_{\mu})L}{(m_{\pi}^2-m_{\mu}^2)^2 E_{\mu}^2 \cos \theta_{\mu} }
\end{eqnarray}
 where
\[ \Delta m^2 \equiv m_1^2-m_2^2 \]      
 The superscripts indicate the particles whose propagators contribute to the 
 interference phase. Also $E_{\pi}$, $E_{\nu}$ and $E_{\mu}$ are the energies of the
 parent $\pi$ and the decay $\nu$ and $\mu$ and $\theta_{\nu}$, $\theta_{\mu}$ the angles
 between the pion and the neutrino, muon flight directions. In Eqns.(1.5) to (1.7) terms
 of order $m_1^4$, $m_2^4$, and higher, are neglected, and in Eqns.(1.6) and (1.7)
 ultrarelativistic kinematics with $E_{\pi,\mu} \gg m_{\pi,\mu}$ is assumed.
 Formulae for the oscillation phase of neutrino oscillations following muon decays
 or nuclear $\beta$-decays at rest, calculated in a similar manner to Eqn(1.5),
 are given in Section 3 below.
 \par A brief comment is now made on the generality and the covariant nature of
  the calculations presented in this paper. Although the fundamental formula
 (1.2) is valid in both relativistic and non-relativistic quantum mechanics,
 it was developed in detail by Feynman~\cite{Feyn2,Feyn3} only for the 
 non-relativistic case. For the conditions of the calculations performed
 in the present paper (propagation of particles in free space) the invariant
 space-time propagator can either be derived (for fermions) from the Dirac
 equation, as originally done by Feynman~\cite{Feyn1} or, more generally,
 from the covariant Feynman path integral for an arbitary massive particle,
 as recently done in Ref.~\cite{Moh1}. In the latter case, the invariant 
 propagator for any stable particle with pole mass $m$, 
 between space-time points $x_i$ and $x_f$ in free space is given
 by the path integral~\cite{Moh1}:
\begin{equation}
 K(x_f,x_i;m) = \int {\cal D}[x(\tau)] \exp \left \{ -\frac{im}{2} \int_{x_i}^{x_f}
  \left( \frac{dx}{d\tau} \cdot \frac{dx}{d\tau}+1 \right)d \tau\right \} 
 \end{equation}
 where $\tau$ is the proper time of the particle. By splitting the 
 integral over $x(\tau)$ on the right side of Eqn(1.8) into the product
 of a series of infinitesimal amplitudes corresponding to small segments,
 $\Delta \tau$, Gaussian integration may be performed over the
 intermediate space-time points. Finally, integrating over the proper time 
 $\tau$, the analytical form of the propagator is found to be~\cite{Moh1}:
 \begin{equation}
 K(x_f,x_i;m) \simeq \left(\frac{m^2}{4 \pi i s}\right) H_1^{(2)}(ims)
 \end{equation}
 where\footnote{The metric for four-vector products is time-like.}:
\[ s =\sqrt{(x_f-x_i)^2}   \]
 and $ H_1^{(2)}$ is a first order Hankel function of the second kind,
 in agreement with Ref.~\cite{Feyn1}.
\par In the asymptotic region where $s \gg m^{-1}$, or for the propagation
 of on-shell particles~\cite{Moh1}, the Hankel function
 reduces to an exponential and yields the configuration space propagator
 $ \simeq \exp(-ims)$ of Eqn(2.11) below. It is also shown in Ref.~\cite{Moh1}
 that energy and momentum is exactly
 conserved in the interactions and decays of all such `asymptotically
 propagating' particles. The use of quasi-classical
 particle trajectories and the requirement of exact energy-momentum
 conservation are crucial ingredients of the calculations presented
 below.
\par The structure of the paper is as follows. In the following Section
 the case of neutrino or muon oscillations following pion decay at rest
 is treated. Full account is taken of the momentum wave-packets of the 
 progagating neutrinos and muons resulting from the Breit-Wigner
 amplitudes describing the distributions of the physical masses of the 
 decaying pion and daughter muon. The corresponding oscillation
 damping corrections and phase shifts are found to be very small, 
 indicating that the quasi-classical (constant velocity) approximation
 used to describe the neutrino and muon trajectories is a very good one.
 The incoherent effects, of random thermal motion of the source pion,
 and of finite source
 and detector sizes, on the oscillation probabilities
 and the oscillation phases, are also calculated. 
 These corrections
 are found to be small in typical experiments, but much 
 larger than those generated by the coherent momentum wave packets.
  In Section 3, formulae are derived to describe neutrino oscillations
  following muon decay at rest or the $\beta$-decay of radioacive 
  nuclei. These are written down by direct analogy with those
  derived in the previous Section for pion decay at rest. In Section 4,
  the case of neutrino and muon oscillations following pion decay
 in flight is treated. In this case the two-dimensional spatial
 geometry of the particle trajectories must be related to the
 decay kinematics of the production process. Due to the non-applicability
  of the ultrarelativistic approximation to the kinematics of the
 muon in the pion rest frame, the calculation, although straightforward,
 is rather tedious and lengthy for the case of muon
 oscillations, so the details are relegated to an appendix. 
  Finally, in Section 5, the positive aspects and shortcomings of 
  previous treatments in the literature of the quantum mechanics
 of neutrino and muon oscillations are discussed in comparison with the
 method and results of the present paper.
    
\SECTION{\bf{Neutrino and muon oscillations following pion decay at rest}}

To understand clearly the different physical hypotheses and approximations underlying
 the calculation of the particle oscillation effects it is convenient to analyse
 a precise experiment. This ideal experiment is, however, very similar to  
 LNSD ~\cite{LNSD} and KARMEN~\cite{KARMEN} except that neutrinos are produced from
 pion, rather than muon, decay at rest. 
 \par The different space-time events that
 must be considered in order to construct the probability amplitudes for the case of
 neutrino oscillations following pion decay at rest are shown in Fig.1. A $\pi^+$ passes
 through the counter
 C$_{\rm A}$, where the time $t_0$ is recorded, and comes to rest in a thin stopping target T
 (Fig.1a)). 
 For simplicity, the case of only two neutrino mass eigenstates $\nu_1$ and $\nu_2$
 of masses $m_1$ and $m_2$ ($m_1$ $>$ $m_2$) is considered. The pion at rest constitutes
 the initial state of the quantum mechanical probability amplitudes. The final state is 
 an $e^-p$ system produced, at time $t_D$, via the process $\nu_e n \rightarrow e^-p$ 
 at a distance $L$ from the decaying $\pi^+$ (Fig.1d)). Two
 different physical processes may produce the observed $e^-p$ final state, as shown in
 Fig.1b) and 1c), where the pion decays either at time $t_1$ into $\nu_1$ or at time
 $t_2$ into $\nu_2$. The probability amplitudes for these processes are, up to an
 arbitary normalisation constant: 
 \begin{eqnarray}
 A_i&=& \int <e^-p|T|n \nu_e><\nu_e|\nu_i>D( x_f-x_i,t_D-t_i, m_i )
 BW(W_{\mu(i)}, m_{\mu}, \Gamma_{\mu})<\nu_i|\nu_\mu>  \nonumber \\ 
 &  & <\nu_\mu \mu^+|T|\pi^+> e^{-\frac{\Gamma_\pi}{2}(t_i-t_0)}D( 0,t_i-t_0, m_\pi )
BW(W_{\pi}, m_{\pi}, \Gamma_{\pi})dW_{\mu(i)}~~i=1,2 
  \end{eqnarray} 
 Note that following the conventional `$fi$' (final,initial) ordering of the indices
 of matrix elements in quantum mechanics, the path amplitude is written from right to
 left in order of increasing time. This ensures also correct matching of `bra' and
 `ket' symbols in the amplitudes. In Eqn(2.1),
  $<e^-p|T|n \nu_e>$, $<\nu_\mu \mu^+|T|\pi^+>$ are the invariant amplitudes of the
   $\nu_e$ charged current scattering and pion decay processes ,
 $BW(W_{\mu(i)}, m_{\mu}$, $\Gamma_{\mu})$ and $BW(W_{\pi}, m_{\pi}$, $\Gamma_{\pi})$
 are relativistic Breit-Wigner amplitudes, $<\nu_e|\nu_i>$ and  $<\nu_i|\nu_\mu>$ are 
 amplitudes describing the mixture of the flavour and mass neutrino eigenstates, and $D$ 
 is the Lorentz-invariant configuration space propagator~\cite{Moh1,Feyn1} of
 the pion or neutrino.
 The pole masses and total decay widths of the pion and muon
  are denoted by $m_{\pi}$, $\Gamma_{\pi}$ and $m_{\mu}$, $\Gamma_{\mu}$ respectively.
  For simplicity, phase space factors accounting for different observed final states
  are omitted in Eqn(2.1) and subsequent formulae. 
 \par Because the amplitudes and propagators in Eqn(2.1) are calculated using relativistic
  quantum field theory, and the neutrinos propagate over macroscopic distances, it is a 
  good approximation, as already discussed in the previous section, to assume exact
  energy-momentum conservation in the pion decay 
  process, and that the neutrinos are on their mass shells, {\it i.e.} $p_i^2=m_i^2$, where
  $p_i$ is the neutrino energy-momentum four-vector. In these circumstances the 
  neutrino propagators correspond to classical, rectilinear, particle trajectories. 
  \par The pion and muon are unstable particles whose physical masses $W_{\pi}$ and
   $W_{\mu(i)}$ differ from the pole masses $ m_{\pi}$ and $m_{\mu}$ appearing in the
   Breit-Wigner amplitudes and covariant space-time propagators in Eqn(2.1). The 
   neutrino momentum $P_i$ will depend on these physical masses according to the
   relation: 
 \begin{equation}
P_i =\frac{ \left[[W_{\pi}^2-(m_i+W_{\mu(i)})^2][W_{\pi}^2-(m_i-W_{\mu(i)})^2]
\right]^{\frac{1}{2}}}
       {2 W_{\pi}}
\end{equation}
 
 Note that, because the initial state pion is the same in the two path amplitudes in
 Eqn(2.1) $W_{\pi}$ does not depend on the neutrino mass index $i$. However, since
 the pion decays resulting in the production of $\nu_1$ and $\nu_2$ are independent
 physical processes, the physical masses of the unobserved muons, $W_{\mu(i)},~~i=1,2$, recoiling 
 against the two neutrino mass eigenstates are not, in general, the same. In the
 following kinematical calculations sufficient accuracy is achieved by retaining
 only quadratic terms in the neutrino masses, and terms linear in the small 
 quantities : $\delta_{\pi} = W_{\pi}-m_{\pi}$, $\delta_i =W_{\mu(i)}-m_{\mu}$.
 This allows simplification of the relativistic Breit-Wigner amplitudes:
 \begin{eqnarray} 
 BW(W,m,\Gamma) &\equiv& \frac{\Gamma m}{W^2-m^2+im\Gamma} \nonumber \\
  &=& \frac{\Gamma m}{\delta(2m+\delta)+im\Gamma} \nonumber \\
  &=& \frac{\Gamma }{2(\delta+i\frac{\Gamma}{2})}~+~ O(\delta^2) \nonumber \\
  &\equiv&  BW(\delta,\Gamma)~+~ O(\delta^2) 
  \end{eqnarray}
 Developing Eqn(2.2) up to first order in $m_i^2$, $\delta_i$ and $\delta_{\pi}$
 yields the relation :
 \begin{eqnarray}
P_i & = & P_0\left[1-\frac{m_i^2(m_{\pi}^2+m_{\mu}^2)}{(m_{\pi}^2-m_{\mu}^2)^2}
 +\frac{\delta_{\pi}}{m_{\pi}}\frac{(m_{\pi}^2+m_{\mu}^2)}{(m_{\pi}^2-m_{\mu}^2)} 
 \right. \nonumber \\
 &   & \left.
 -\frac{2\delta_i m_{\mu}}{m_{\pi}^2-m_{\mu}^2}
 +\frac{\delta_{\pi} m_i^2(m_{\pi}^2+m_{\mu}^2)}{m_{\pi}(m_{\pi}^2-m_{\mu}^2)^2}
\right]
 \end{eqnarray}
where 
 \begin{equation}
 P_0 = \frac{m_{\pi}^2-m_{\mu}^2}{2 m_{\pi}}~ = ~ 29.8 \rm{MeV}
\end{equation}
 The term $\simeq \delta_{\pi} m_i^2$ which is also included in Eqn(2.4)
 gives a negligible O($m_i^4$) contribution to the neutrino oscillation 
 formula. For muon oscillations, however, it gives a term of O($m_i^2$)
 in the interference term, as discussed below. 
 Similarly, the exact formula for the neutrino energy:
 \begin{equation}
E_i = \frac{W_{\pi}^2-W_{\mu(i)}^2+m_i^2}{2 W_{\pi}}
\end{equation}
 in combination with Eqn(2.4) gives, for the neutrino velocity:
\begin{equation}
v_i = \frac{P_i}{E_i}=1-\frac{m_i^2}{2 P_0^2}\left[1-
\frac{2\delta_{\pi} (m_{\pi}^2+m_{\mu}^2)}{ m_{\pi} (m_{\pi}^2-m_{\mu}^2)}
+\frac{4\delta_i m_{\mu}}{m_{\pi}^2-m_{\mu}^2}\right]~+~O(m_i^4,\delta_{\pi}^2,\delta_i^2)
\end{equation}
This formula will be used below to calculate the neutrino times-of-flight:
 $t_i^{fl}~~i=1,2$. 
\par The unitary transformation specifying the mixing of the flavour
and mass neutrino eigenstates is defined by a single real mixing angle
$\theta$:
\begin{eqnarray}
|\nu_e>& = &\cos\theta|\nu_1>+\sin\theta|\nu_2> \\
|\nu_\mu>& = &-\sin\theta|\nu_1>+\cos\theta|\nu_2> 
\end{eqnarray}
where the flavour eigenstates $|\alpha>,~\alpha=\nu_e,\nu_{\mu}$ and the 
mass eigenstates  $|a>,~a=\nu_1,\nu_2$ respect the orthonormality conditions:
\begin{equation}
<\alpha|\beta>=\delta_{\alpha \beta},~~~<a|b>=\delta_{a b}
\end{equation}
\par The parts of the amplitudes requiring the most careful discussion 
are the invariant space-time propagators $D$, as it is mainly their treatment
that leads to the different result for the neutrino oscillation phase
 found in the present paper, as compared to those having
 previously appeared in the literature. In the limit of large time-like
 separations, the propagator may be written
 as~\cite{Moh1,Feyn1}:
\begin{equation}
D(\Delta x,\Delta t,m)=\left(\frac{m}{2\pi i \sqrt{(\Delta t)^2-(\Delta x)^2}}\right)^{\frac{3}{2}}
\exp[-im \sqrt{(\Delta t)^2-(\Delta x)^2}] 
\end{equation}
 $D$ is the amplitude for a particle, originally at a space-time point ($\vec{x}_i$, $t_i$),
 to be found at ($\vec{x}_f$, $t_f$) and $\Delta\vec{x} \equiv \vec{x}_f-\vec{x}_i$, 
 $\Delta t \equiv t_f-t_i$. In the following, according to the geometry of the experiment 
 shown in Fig.1, only one spatial coordinate will be considered ($\Delta x\equiv x_f-x_i$)
and only the exponential factor in Eqn(2.11), containing the essential phase information
 for particle oscillations will be retained in the amplitudes. Solid angle correction factors,
taken into account by the factor in large brackets in Eqn(2.11), are 
 here neglected, but are easily included in the final oscillation formulae.  
Writing then
\begin{equation}
D(\Delta x,\Delta t,m) \simeq  \exp[-im \sqrt{(\Delta t)^2-(\Delta x)^2}] = \exp[-im\Delta \tau]
 \equiv \exp[-i\Delta \phi] 
\end{equation}
it can be seen that the increment in phase of the propagator, $\Delta \phi$, when the particle
 undergoes the space-time displacement ($\Delta x$, $\Delta t$) is a Lorentz invariant quantity
 equal to the product of the particle mass and the increment, $\Delta \tau$, of proper time.
 Using the relativistic time dilatation formula:
\begin{equation}
 \Delta t = \gamma \Delta \tau = \frac{E}{m} \Delta \tau
\end{equation}
 and also the relation, corresponding to a classical, rectilinear, particle trajectory:
\begin{equation}
 \Delta t = \frac{L}{v} = \frac{E}{p} L
\end{equation}  
gives, for the phase increments corresponding to the paths of the neutrinos and the pion
in Fig.1:
\begin{eqnarray}
\Delta \phi_i^{\nu}& = & m_i \Delta \tau_i =\frac{m_i^2}{E_i}
 \Delta t_i = \frac{m_i^2}{P_i}L \nonumber  \\
 & = & \frac{m_i^2 L}{P_0}\left[1 -
\frac{\delta_{\pi}}{m_{\pi}}\frac{(m_{\pi}^2+m_{\mu}^2)}{(m_{\pi}^2-m_{\mu}^2)} 
 +\frac{2\delta_i m_{\mu}}{m_{\pi}^2-m_{\mu}^2}\right]  \\
 \Delta \phi_i^{\pi}& = & m_{\pi} (t_i-t_0)
  =  m_{\pi} (t_D-t_0)-\frac{m_{\pi}L}{v_i}  \nonumber \\
 & = & m_{\pi} (t_D-t_0)-m_{\pi}L\left\{1+\frac{m_i^2}{2 P_0^2}\left[1-
\frac{2\delta_{\pi}}{m_{\pi}}\frac{(m_{\pi}^2+m_{\mu}^2)}{(m_{\pi}^2-m_{\mu}^2)} 
+\frac{4\delta_i m_{\mu}}{m_{\pi}^2-m_{\mu}^2}\right]\right\} 
\end{eqnarray}
\par Making the substitution $t_i-t_0 \rightarrow t_D-t_0-L/v_i$ in the
exponential damping factor due to the pion lifetime in Eqn(2.1) and using 
 Eqns(2.3),(2.12),(2.15) and (2.16) Eqns(2.1) may be written as:
 \begin{eqnarray}
 A_i & = & \int <e^-p|T|n \nu_e><\nu_e|\nu_i><\nu_i|\nu_\mu><\nu_\mu \mu^+|T|\pi^+>
  \frac{\Gamma_{\mu}}{2} \frac{e^{i\alpha_i \delta_i}}{( \delta_i+i\frac{\Gamma_{\mu}}{2})} \nonumber \\
 &   &  \frac{\Gamma_{\pi}}{2} \frac{e^{i\alpha_{\pi}(i)
 \delta_{\pi}}}{(\delta_{\pi}+i\frac{\Gamma_{\pi}}{2})}
 e^{ i\phi_0-\frac{\Gamma_\pi}{2}(t_D-t_0-t_i^{fl}) }
\exp i \left[ \frac{ m_i^2}{P_0}\left(
 \frac{m_{\pi}}{2 P_0}-1\right)L \right] d \delta_i~~~i=1,2 
 \end{eqnarray} 
where
\begin{equation}
\phi_0 \equiv m_{\pi}(L-t_D+t_0)
\end{equation}
\begin{equation}
\alpha_i \equiv \frac{ 4 m_i^2 m_{\mu} m_{\pi} ( m_{\pi}^2+ m_{\mu}^2) L}
 {( m_{\pi}^2- m_{\mu}^2)^3}\left[1-\frac{i\Gamma_{\pi} m_{\pi}}{m_{\pi}^2+ m_{\mu}^2}\right]  
\end{equation}
\begin{equation}
\alpha_{\pi}(i) \equiv -\frac{ 2 m_i^2  ( m_{\pi}^2+ m_{\mu}^2)^2 L}
 {( m_{\pi}^2- m_{\mu}^2)^3}\left[1-\frac{i\Gamma_{\pi} m_{\pi}}{m_{\pi}^2+ m_{\mu}^2}\right]  
\end{equation}
and 
\begin{equation}
t_i^{fl} = L(1+\frac{m_i^2}{2 P_0^2})+O(m_i^4)
\end{equation}

\par To perform the integral over $\delta_i$ in Eqn(2.17) it is convenient to 
 approximate the modulus squared of the Breit-Wigner amplitude by 
 a Gaussian, {\it via} the substitution:
\begin{equation}
\frac{\frac{\Gamma}{2}}{\delta+i\frac{\Gamma}{2}} = {\frac{\Gamma}{2}}\left(
\frac{\delta-i\frac{\Gamma}{2}}{\delta^2+\frac{\Gamma^2}{4}}\right) \rightarrow 
\frac{2}{\Gamma}(\delta-i\frac{\Gamma}{2})\exp\left(-\frac{3 \delta^2}{\Gamma^2}\right)
\end{equation}
where the width of the Gaussian is chosen so that it has approximately 
 the same full width at half maximum, $\Gamma$, as the Breit-Wigner function.
After the substitution (2.22), the integral over $\delta_i$ in Eqn(2.17) is 
 easily evaluated by a change of variable to `complete the square' in the
 argument of the exponential, with the result:
\begin{equation}
I_i =\frac{2}{\Gamma_{\mu}}
 \int_{-\infty}^{\infty}(\delta_i-i\frac{\Gamma_{\mu}}{2}))\exp\left(-\frac{3 \delta_i^2}
{\Gamma_{\mu}^2}+i\alpha_i \delta_i\right)d \delta_i  = i\sqrt{\frac{\pi}{3}}\Gamma_{\mu}
\exp\left(-\frac{\alpha_i^2
 \Gamma_{\mu}^2}{12}
\right)
 \left[\frac{\alpha_i \Gamma_{\mu}}{3}-1\right]
\end{equation}
 Since $\Gamma_{\pi} = 2.53 \times 10^{-14}$ MeV, the term dependent on
 $\Gamma_{\pi}$ in Eqn(2.19) may be neglected, so that $\alpha_i$ may be taken as 
 a real number with the numerical value: 
\begin{equation}
\alpha_i = 3.1 \times 10^{14}\left(\frac{m_i}{m_{\pi}}\right)^2 L(m)~~\rm{MeV}^{-1}
\end{equation}
 For typical physically interesting values (see below) of 
$m_i = 1$ eV and $L = 30$ m, $\alpha_i$ takes the value 0.48 MeV$^{-1}$, so that
\[ \alpha_i \Gamma_{\mu} = 0.48 \times 3.00\times 10^{-16} = 1.44 \times 10^{-16} \]  
Then, to very good accuracy, $I_1 = I_2 = -i\sqrt{\pi/3} \Gamma_{\mu}$, independently of
 the neutrino mass. It follows that for neutrino oscillations, the muon mass dependence
 of the amplitudes may be neglected for any physically interesting values of $m_i$ and $L$.
 \par From Eqns (2.17) and (2.23) the probability to observe the reaction
 $\nu_e n \rightarrow e^-p$ at distance $L$ from the pion decay point and at time $t_D$ is:      
\begin{eqnarray}
P(e^-p|L,t_D) & = & |A_1+A_2|^2 \nonumber  \\
              & = & \frac{\pi  \Gamma_{\mu}^2}{3}|<e^-p|T|n\nu_e>|^2 
 |<\nu_{\mu} \mu^+|T|\pi^+>|^2    \nonumber \\
  &  & \times  \sin^2\theta \cos^2\theta
e^{-\Gamma_{\pi}(t_D-t_0)} \frac{ \Gamma_{\pi}^2}{4\left(\delta_{\pi}^2+\frac{\Gamma_{\pi}^2}{4}\right)}   \\
  &  & \times \left\{ e^{\Gamma_{\pi} t_1^{fl}}+ e^{\Gamma_{\pi} t_2^{fl}}\right. \nonumber \\
  &  & \left. - 2  e^{\Gamma_{\pi}\frac{(t_1^{fl}+t_2^{fl})}{2}}
{\it Re} \exp i \left[\frac{\Delta m^2}{P_0}\left(
 \frac{m_{\pi}}{2 P_0}-1\right)L+[\alpha_{\pi}(1)-\alpha_{\pi}(2)]\delta_{\pi}\right]\right\} \nonumber
\end{eqnarray}
  The time dependent exponential
 factors in the curly brackets of Eqn(2.25) are easily understood. If $m_1>m_2$ then
 $t_1^{fl} > t_2^{fl}$. This implies that the neutrino of mass $m_1$ results from an earlier
 decay than the neutrino of mass $m_2$, in order to be detected at the same time. Because of
 the exponential decrease with time of the pion decay amplitude, the contribution to the
 probability of the squared amplitude for the neutrino of mass $m_1$ is larger. The 
 interference term resulting from the product of the decay amplitudes of the two 
 neutrinos of different mass, has an exponential factor that is the harmonic mean of
 those of the squared amplitudes for  each neutrino mass eigenstate, and so is
 also suppressed relative to the squared amplitude for the neutrino of mass $m_1$.
 The integral over the physical pion mass is readily performed by replacing the
 Breit-Wigner function by a Gaussian as in Eqn(2.22). This leads to an overall
 multiplicative constant $\sqrt{\pi/3} \Gamma_{\pi}$ and a factor:
  \[ \exp [-(\alpha_{\pi}(1)-\alpha_{\pi}(2))^2 \Gamma_{\pi}^2 /12] \] multiplying the
 interference term. For $\Delta m^2= (1\rm{eV})^2$ and $L = 30$m the numerical 
 value of this factor is $\exp(-1.3 \times 10^{-29})$. This tiny correction is 
 neglected in the following equations.
\par Integrating over $t_D$ gives the average probability to observe the $e^-p$ final 
 state at distance $L$:
\begin{eqnarray}
P(e^-p|L) & = & \frac{\pi^{\frac{3}{2}} \Gamma_{\mu}^2}{3\sqrt{3}} |<e^-p|T|n\nu_e>|^2
 |<\nu_{\mu} \mu^+|T|\pi^+>|^2
 \sin^2\theta \cos^2\theta \nonumber \\
  &  & \times \left\{1-\exp[-\frac{\Gamma_{\pi} m_{\pi}^2 \Delta m^2}{( m_{\pi}^2- m_{\mu}^2)^2}L]
\cos\frac{2 m_{\pi} m_{\mu}^2 \Delta m^2}{( m_{\pi}^2- m_{\mu}^2)^2}L\right\} 
\end{eqnarray}
Where all kinematical quantities are expressed in terms of $\Delta m^2$, $m_{\pi}$ and 
 $m_{\mu}$.  
Note that the minimum value of $t_D$ is $t_0+t_1^{fl}$,  $t_0+t_2^{fl}$
and $t_0+t_1^{fl}$ for the squared amplitude terms of neutrinos of mass $m_1$, $m_2$
and the interference term, respectively. On integrating over $t_D$, the squared amplitude
terms give equal contributions, the larger amplitude for mass $m_1$ being exactly compensated
 by a smaller range of integration. The exponential damping factor in the interference term in
 Eqn(2.26) is derived using the relations:
\begin{equation}
t_1^{fl}-t_2^{fl} = \left(\frac{1}{v_1(\nu)}-\frac{1}{v_2(\nu)}\right)L
 \simeq (v_2(\nu)-v_1(\nu))L
\end{equation}
and  
\begin{equation}
 v_i(\nu)=1-\frac{m_i^2}{2 P_0^2}+O(m_i^4)~~~~i=1,2
\end{equation}
to obtain
\begin{equation}
t_1^{fl}-t_2^{fl} = \frac{(m_1^2-m_2^2)L}{2 P_0^2} = \frac{\Delta m^2 L}{2 P_0^2}
\end{equation}
 The damping factor arises because the difference in the  times-of-flight of the two
 neutrino paths is limited by the pion lifetime. It will be seen below, however, that
 for distances $L$ of practical interest for the observation of neutrino oscillations,
 the damping effect is tiny.
 \par The part of the oscillation phase in Eqn(2.25) originating from the neutrino
 propagators (the term associated with the `1' within the large curved brackets) differs
 by a factor two from
 the corresponding expression in the standard formula. 
 The contribution to the oscillation phase of the propagator
 of the decaying pion ( the term associated with $m_{\pi}/(2 P_0)$ in the large curved
 brackets of Eqn(2.24))  
 has not been taken into account in any published calculation known to the author of
 the present paper.
 The oscillation phase in Eqn(2.26) is $2 m_{\mu}^2/( m_{\pi}^2- m_{\mu}^2) = 2.685$ times
 larger than that given by the standard formula (1.1). 
 For $L=30$m, as in the LNSD
 experiment, the first oscillation maximum occurs for $\Delta m^2 = 0.46 (eV)^2$. 
 Denoting by $\phi_{12}^{\nu,\pi}$ the phase of the cosine interference term in Eqn(2.26),
 the  pion lifetime damping factor can be written as:
\begin{equation}
F^{\nu}(\Gamma_{\pi})=\exp\left(-\Gamma_{\pi}\frac{m_{\pi}}{2 m_{\mu}^2}
 \phi_{12}^{\nu,\pi}\right) = \exp(-1.58 \times 10^{-16}\phi_{12}^{\nu,\pi}) 
\end{equation}
so the damping effect is vanishingly small when $\phi_{12}^{\nu,\pi} \simeq 1$.
\par The oscillation formula (2.26) is calculated on the assumption that the
 decaying pion is at rest at the precisely defined position $x_i$. In fact,
 the positive pion does not bind with the atoms of the target, but will
 rather undergo random thermal motion. This has three effects: an uncertainy
 in the value of $x_i$, a Doppler shift of the neutrino energy and a time dilatation
 correction correction factor of $1/\gamma_{\pi}$ in the equation (2.16) for the pion
 phase increment. 
 Assuming that the target is at room temperature (T= $270^{\circ}$ K), the mean
 kinetic energy of $3kT/2$ correponds to a mean pion momentum of
 $2.6 \times 10^{-3}$ MeV and a mean velocity of $\simeq 5.6$ km/sec. The pion will
 move, in one mean lifetime ($2.6 \times 10^{-8}$ sec), a distance of 146$\mu m$.
 This is negligible as compared to $L$ (typically $\ge$ 30m) and so Eqn(2.26)
 requires no modification to account for this effect. 
 \par The correction factor due to the Doppler effect
 and time dilatation is readily calculated on the assumption
 of a Maxwell-Boltzmann distribution of the pion momentum:
\begin{equation}
\frac{dN}{dp_{\pi}} \simeq p_{\pi}^2 \exp\left( -\frac{ p_{\pi}^2}{\overline{p}_{\pi}^2}\right)
\end{equation}
Here $\overline{p}_{\pi} = \sqrt{2kTm_{\pi}} = 2.64 \times 10^{-3}$ MeV. Details 
 of the calculation are given in Appendix A. The interference term in Eqn(2.26) is
 modified by a damping factor:
\begin{equation}
F^{\nu}(Dop) = \exp \left\{ -\left(\frac{\overline{p}_{\pi}\Delta m^2}
{2 m_{\pi} P_0}\left[\frac{m_{\pi}}{P_0}-1\right]L\right)^2 \right\}
\end{equation}
 while the argument of the cosine term acquires an additional phase factor:
\begin{equation}
\phi^{\nu}(Dop) = \frac{3}{4}\left(\frac{\overline{p}_{\pi}}{m_{\pi}}
\right)^2 \frac{\Delta m^2}{P_0}\left[\frac{3 m_{\pi}}{2 P_0}-1 \right]L 
\end{equation}
For  $\phi_{12}^{\nu,\pi} =  1$, $F^{\nu}(Dop) = 1-6.7\times10^{-10}$ and
$\phi^{\nu}(Dop) = 1.2\times10^{-9}$.
\par If the target in which the pion stops is of thickness $\ell_T$,
 then the effect of different stopping points of the $\pi$ (assumed uniformly
 distributed) is to multiply the interference term in (2.26) by the factor:
\begin{equation}
F_{Targ}^{\nu} = \frac{(m_{\pi}^2-m_{\mu}^2)^2}{m_{\pi}m_{\mu}^2\Delta m^2\ell_T}
\sin\left(\frac{m_{\pi}m_{\mu}^2\Delta m^2\ell_T}{(m_{\pi}^2-m_{\mu}^2)^2}\right)
\end{equation} 
 If the position of the neutrino interaction point within the target has an 
 uncertainy of $\pm \ell_D/2$ a similar correction factor is found, with the
 replacement $\ell_T \rightarrow \ell_D$  in Eqn(2.34). The calculation 
 of this correction factor is also described in Appendix A.
 
\par If the target T in which the pion comes to rest (Fig.1a)) is chosen to
 be sufficiently thin, the pion decay process may be detected by observing 
 the recoil muon in the counter C$_{\rm A}$ at times $t_1$ or $t_2$  (Fig.1b) 
 or 1c)). A sufficiently accurate measurement of the times  $t_1$, $t_2$
 and $t_D$ would, in principle, enable separation of the different processes in Figs.1b) and
 1c) by the observation of separated peaks in the time-of-flight distribution
 at $t_1^f=t_D-t_1$ and $t_2^f=t_D-t_2$. In this case the interference term in
 Eqn(2.26) vanishes as the two alternative space-time paths leading to the 
 neutrino interaction are distinguishable. However, for $L=30$m and 
 $\Delta m^2 = 1$(eV)$^2$
 the difference in the times of flight is only $5.6 \times 10^{-23}$ sec, more than ten
orders of magnitude smaller than can be measured with existing techniques.
 As discussed in Section 5 below, the momentum smearing due to the Doppler
 effect at room temperature is some eleven orders of magnitude larger than the
  shift due to a neutrino mass difference with $\Delta m^2 = 1$(eV)$^2$. Thus, 
 even with infinitely good time resolution, separation of such neutrino mass eigenstates
 by time-of-flight is not possible. 
 
\par The ideal experiment, described above to study neutrino oscillations, is easily
 adapted to the case of oscillations in the decay probability of muons produced
 by charged pion decay at rest. As previously pointed out in Ref.~\cite{SWS},
 such oscillations will occur if neutrinos with different
 masses exist. As before, the pion stops in the target T at time $t_0$
 (see Fig.2a)). At time $t_1$ the pion decays into $\nu_1$ and the corresponding recoil
 muon ($\mu_1$), whose passage is recorded in the counter C$_{\rm B}$ (Fig.2b)). Similarly, a 
 decay into $\nu_2$ and $\mu_2$ may occur at time $t_2$ (Fig.2c)). With a suitable
 choice of the times $t_1$ and $t_2$, such that muons following the alternative
 paths both arrive at the same time $t_D$ at the point $x_f$, interference 
 occurs between the path amplitudes when muon decay occurs at the space-time
 point ($x_f$, $t_D$) in the detector D (Fig.2d)). The path amplitudes corresponding
 to muons
 recoiling against neutrinos of mass $m_1$ and $m_2$ are:
 \begin{eqnarray}
 A_i & = & \int <e^+\nu_e\overline{\nu}_{\mu}|T|\mu^+>
 e^{-\frac{\Gamma_{\mu}v^{\mu}_i}{2\gamma^{\mu}_i}L}D( x_f-x_i,t_D-t_i, m_{\mu} )
 BW(W_{\mu(i)}, m_{\mu}, \Gamma_{\mu})<\nu_i|\nu_\mu>  \nonumber \\ 
 &  & <\nu_\mu \mu^+|T|\pi^+> e^{-\frac{\Gamma_\pi}{2}(t_i-t_0)}D(0,t_i-t_0, m_\pi )
 BW(W_{\pi}, m_{\pi}, \Gamma_{\pi}) dW_{\mu(i)}~~i=1,2 
 \end{eqnarray} 
 The various factors in these equations are defined, {\it mutadis mutandi}, as in Eqn(2.1).
 
 \par With the same approximations, concerning the neutrino masses and the physical
 pion and muon masses, as those made above, the velocity of the muon recoiling against
 the neutrino mass eigenstate $\nu_i$ is:
 \begin{eqnarray}
v^{\mu}_i& = &v^{\mu}_0\left[1-
\frac{4 m_i^2 m_{\pi}^2 m_{\mu}^2}{(m_{\pi}^2-m_{\mu}^2)^2(m_{\pi}^2+m_{\mu}^2) }
+\frac{4\delta_{\pi}m_{\pi} m_{\mu}^2}{m_{\pi}^4-m_{\mu}^4} \right. \nonumber \\
 &  & \left.
-\frac{4\delta_i m_{\mu} m_{\pi}^2 }{m_{\pi}^4-m_{\mu}^4}
-\frac{8 \delta_{\pi} m_i^2 m_{\pi}^3 m_{\mu}^2
 (3m_{\mu}^2-m_{\pi}^2)}{(m_{\pi}^4-m_{\mu}^4)^2(m_{\pi}^2-m_{\mu}^2) } \right]
\end{eqnarray}
where 
\begin{equation}
v^{\mu}_0 = \frac{m_{\pi}^2-m_{\mu}^2}{m_{\pi}^2+m_{\mu}^2}
\end{equation} 
 Comparing with Eqn(2.7), it can be seen that for the muon case,
 unlike that where neutrino interactions are observed, there are pion and muon mass
 dependent correction terms
 that are
 independent of the neutrino masses, implying a velocity smearing effect due to 
 the physical pion and muon masses that is $\simeq m_{\pi}^2/m_i^2 $ larger than for
 the case of neutrino oscillations.
 
 \par The phase increments corresponding to the paths of the muons and the pion in Fig.~2 
 are, using (2.4)\footnote{Note that, in the pion rest frame $P_i = P^{\mu}_i$.} and (2.12)-(2.14):
\begin{eqnarray} 
\Delta \phi^{\mu}_i & = & \frac{m_{\mu}^2 L}{P_i^{\mu}} = \frac{m_{\mu}^2 L }{P_0}\left[
 1+\frac{m_i^2 E_0^{\mu}}{2 m_{\pi} P_0^2}
-\frac{\delta_{\pi}}{m_{\pi}}\frac{(m_{\pi}^2+m_{\mu}^2)}{(m_{\pi}^2-m_{\mu}^2)} 
 +\frac{2\delta_i m_{\mu}}{m_{\pi}^2-m_{\mu}^2} \right.   \nonumber \\
 &  & \left. -\frac{\delta_{\pi} m_i^2}{m_{\pi}}\frac{(m_{\pi}^2+m_{\mu}^2)}
{(m_{\pi}^2-m_{\mu}^2)^2} \right]   
\end{eqnarray}
\[ \Delta \phi_i^{\pi(\mu)}  =  m_{\pi} (t_i-t_0)
  =  m_{\pi} (t_D-t_0)-\frac{m_{\pi}L}{v^{\mu}_i} \]
\[ =  m_{\pi} (t_D-t_0) -\frac{m_{\pi}L}{v^{\mu}_0}\left[1+\frac{4 m_i^2 m_{\pi}^2 m_{\mu}^2}
{(m_{\pi}^2-m_{\mu}^2)^2(m_{\pi}^2+m_{\mu}^2)} \right. \]  
\begin{equation}
 \left. -\frac{4 \delta_{\pi} m_{\pi} m_{\mu}^2}{m_{\pi}^4-m_{\mu}^4}
 +\frac{4 \delta_i m_{\mu} m_{\pi}^2 }{m_{\pi}^4-m_{\mu}^4}
 +\frac{8 \delta_{\pi} m_i^2 m_{\pi}^3 m_{\mu}^2 (3 m_{\mu}^2-m_{\pi}^2)}
{(m_{\pi}^4-m_{\mu}^4)^2(m_{\pi}^2-m_{\mu}^2)}\right] 
\end{equation}
 where
\begin{equation}
E_0^{\mu} = \frac{m_{\pi}^2+m_{\mu}^2}{2 m_{\pi}}
\end{equation}  
 \par Using Eqns(2.12),(2.38) and (2.39) to re-write the space-time propagators in
 Eqn(2.35), as well as  Eqn(2.3) for the Breit-Wigner amplitudes gives:
 \begin{eqnarray}
 A_i^{\mu}&=&\int <e^+ \nu_e\overline{\nu}_{\mu}|T|\mu^+> e^{-\frac{\Gamma_{\mu} v_0^{\mu}L}
{2 \gamma_0^{\mu}}}
 <\nu_i|\nu_\mu><\nu_\mu \mu^+|T|\pi^+> \frac{\Gamma_{\mu}}{2} \frac{e^{i\alpha^{\mu} \delta_i}}
{(\delta_i+i\frac{\Gamma_{\mu}}{2})} \frac{\Gamma_{\pi}}{2} \nonumber \\
 &   &  \frac{e^{i\alpha^{\mu}_{\pi}(i)
 \delta_{\pi}}}{(\delta_{\pi}+i\frac{\Gamma_{\pi}}{2})}
 e^{ i\phi^{\mu}_0-\frac{\Gamma_{\pi}}{2}(t_D-t_0-t_{\mu(i)}^{fl})}
\exp i \left[ \frac{ m_{\mu}^2 m_i^2}{2 P^3_0}\left(1-
 \frac{ E_0^{\mu}}{m_{\pi}} \right)L \right] d \delta_i~~i=1,2 
 \end{eqnarray}  
where  
\begin{equation}
\phi^{\mu}_0 \equiv m_{\pi}(\frac{L}{v^{\mu}_0}-t_D+t_0)-\frac{m_{\mu}^2 L}{P_0}
\end{equation}
\begin{equation}
\alpha^{\mu} \equiv \frac{ 4 m_{\mu} m_{\pi} L}
 { m_{\pi}^2- m_{\mu}^2}\left[1-\frac{i\Gamma_{\pi} m_{\pi}}{2(m_{\pi}^2- m_{\mu}^2)}\right]  
\end{equation}
\begin{equation}
\alpha^{\mu}_{\pi}(i) \equiv -\frac{ 2  m_{\mu}^2  L}
 { m_{\pi}^2- m_{\mu}^2}\left[1+
\frac{ m_i^2 (5 m_{\pi}^6-10 m_{\pi}^4 m_{\mu}^2- m_{\pi}^2 m_{\mu}^4 - m_{\mu}^6)}
{(m_{\pi}^4-m_{\mu}^4)(m_{\pi}^2-m_{\mu}^2)^2} 
-\frac{i\Gamma_{\pi} m_{\pi}}{m_{\pi}^2- m_{\mu}^2}\right]  
\end{equation}
and
\begin{equation}
t_{\mu(i)}^{fl} = L\left(\frac{1}{v^{\mu}_0}+\frac{m_i^2 m_{\mu}^2}
{2 m_{\pi} P_0^3}\right)
\end{equation}
Making the substitution (2.22) and performing the integral over $\delta_i$ according
 to Eqn(2.23), the following formula is found for the\ probability for muon decay 
 at distance $L$ and time $t_D$.
\begin{eqnarray}
P(e^+\nu_e\overline{\nu}_{\mu}|L,t_D) & = & |A_1^{\mu}
+A_2^{\mu}|^2 \nonumber  \\
              & = & \frac{\pi  \Gamma_{\mu}^2}{3}
 e^{-\frac{(\alpha^{\mu} \Gamma_{\mu})^2}{6}}[1-\frac{\alpha^{\mu} \Gamma_{\mu}}{3}]^2
 |<e^+\nu_e\overline{\nu}_{\mu}|T|\mu^+>|^2 e^{-\frac{\Gamma_{\mu}v_0^{\mu}}{\gamma_0^{\mu}}L}
 \nonumber \\
    &  &    |<\nu_{\mu} \mu^+|T|\pi^+>|^2 e^{-\Gamma_{\pi}(t_D-t_0)}
  \frac{\Gamma_{\pi}^2}{4(\delta_{\pi}^2+\frac{\Gamma_{\pi}^2}{4})}  \nonumber  \\
  &  &\left\{\sin^2\theta e^{\Gamma_{\pi} t_{\mu(1)}^f} + \cos^2 \theta e^{\Gamma_{\pi}
  t_{\mu(2)}^f}  \right. \nonumber \\
  &  & -2\sin \theta \cos \theta  e^{\frac{\Gamma_{\pi}}{2}(t_{\mu(1)}^f +t_{\mu(2)}^f )}
 \nonumber \\
  &  & \left. {\it Re} \exp i \left[ 
 \frac{m_{\mu}^2 \Delta m^2}{2 P_0^3}\left(1-\frac{ E_0^{\mu}}{m_{\pi}}\right)L
  +[\alpha_{\pi}^{\mu}(1)-\alpha_{\pi}^{\mu}(2)]\delta_{\pi} \right] \right\}
\end{eqnarray} 
  The muon path difference yields the term associated with $E^{\mu}_0/m_{\pi}$ in the 
 interference phase in Eqn(2.46) while the pion path is associated with  `1' in the
 large round
 brackets.  
 The numerical value of the damping factor:
  \[ \exp[-\frac{(\alpha^{\mu} \Gamma_{\mu})^2}{6}]
[1-\frac{\alpha^{\mu} \Gamma_{\mu}}{3}]^2 \]
  resulting from the integral over the physical muon
 mass is, for $L = 30$m, 0.774, so, unlike for the case of neutrino oscillations, the
 correction is by no means negligible. This is because, in the muon oscillation case,
 the leading term of $\alpha^{\mu}$ is not proportional to the neutrino mass squared.
 The non-leading terms proportional to $m_i^2$ have been neglected in Eqn(2.43).
 This correction however effects only the overall normalisation of the oscillation 
 formula, not the functional dependence on $L$ arising from the interference term.
 Integrating over $\delta_{\pi}$ using Eqns(2.22) and (2.23), as well as over $t_D$, gives 
 the probability of muon decay at distance $L$ from the production point, where all
 kinematical quantities are expressed in terms of $\Delta m^2$, $m_{\pi}$ and $m_{\mu}$:
\begin{eqnarray}
P(e^+\nu_e\overline{\nu}_{\mu}|L) & = & \frac{\pi^{\frac{3}{2}} \Gamma_{\mu}^2}
 {3\sqrt{3}} \exp\left[-\frac{8}{3}\left(\frac{\Gamma_{\mu} m_{\mu} m_{\pi}}
 {m_{\pi}^2- m_{\mu}^2}L\right)^2 \right][1- \frac{4}{3} \frac{\Gamma_{\mu} m_{\mu} m_{\pi}}
 { m_{\pi}^2- m_{\mu}^2}L]^2 \nonumber \\
 &  & |<e^+\nu_e\overline{\nu}_{\mu}|T|\mu^+>|^2 \exp \left[-\frac{2 \Gamma_{\mu} m_{\pi} m_{\mu}
(m_{\pi}^2- m_{\mu}^2)}{(m_{\pi}^2+ m_{\mu}^2)^3} L \right]|<\nu_{\mu} \mu^+|T|\pi^+>|^2
 \nonumber \\
  &  &  \left\{1-\sin 2 \theta\exp[-\frac{2 \Gamma_{\pi} m_{\pi}^2 m_{\mu}^2 \Delta m^2}
{( m_{\pi}^2- m_{\mu}^2)^3}L]
\cos \frac{2 m_{\pi} m_{\mu}^2 \Delta m^2}{( m_{\pi}^2- m_{\mu}^2)^2} L \right\}
\end{eqnarray}
 In this expression the correction due to the damping factor of the interference term:
 \[ \exp[-(\alpha_{\pi}^{\mu}(1)-\alpha_{\pi}^{\mu}(2))^2 \Gamma_{\pi}^2/12] \] arising from the
 integral over the physical pion mass has been neglected. For $\Delta m^2= (1\rm{eV})^2$
 and $L = 30$m the numerical  value of this factor is $\exp(-3.9 \times 10^{-31})$.
Denoting by $\phi_{12}^{\mu,\pi}$ the argument of the cosine in Eqn(2.47), the exponential
damping factor due to the pion lifetime may be written as:
\begin{equation}
F^{\mu}(\Gamma_{\pi})= \exp \left(-\frac{\Gamma_{\pi}
 m_{\pi}}{(m_{\pi}^2-m_{\pi}^2)} \phi_{12}^{\mu,\pi} \right)
\end{equation}
 For $ \phi_{12}^{\mu,\pi} = 1$, $F^{\mu}(\Gamma_{\pi})-1 = 4.4 \times 10^{-16}$ so, as in the neutrino
 oscillation case, the pion lifetime damping of the interference term is very small.
\par Corrections due to time dilatation and the Doppler effect are calculated in a similar 
 way to the neutrino oscillation case with the results (see Appendix A):
\begin{equation}
F^{\mu}(Dop) = \exp \left\{ -\left(\frac{\overline{p}_{\pi}m_{\mu}^2\Delta m^2 v^{\mu}_0}
{2 m_{\pi} P_0^3}\left[\frac{3}{2}-\frac{E_0^{\mu}}{m_{\pi}}\right]L\right)^2 \right\}
\end{equation}
and 
\begin{equation}
\phi^{\mu}(Dop) = \frac{3}{2}\left(\frac{\overline{p}_{\pi}}{m_{\pi}}
\right)^2  \frac{m_{\mu}^2\Delta m^2}{P_0^3}\left[1-\frac{E_0^{\mu}}{2 m_{\pi}} \right]L 
\end{equation}
As for neutrino oscillations, the corresponding corrections are very small for
 oscillation phases of order unity.
 \par The phase of the cosine in the interference term is the same in
neutrino and muon oscillations, as can be seen by comparing Eqns(2.26) and (2.47). It follows 
that the target or detector size correction (Eqn(2.34)) is the same in 
 both cases. 
 \par Neutrino and muon oscillations from pion decay at rest have an identical
 oscillation phase for given values of $\Delta m^2$ and $L$. In view of the much
 larger event rate that is possible, it is clearly very advantageous in this case
 to observe muons rather than neutrinos. In fact, it it not necessary to observe 
 muon decay, as in the example discussed above. The oscillation formula applies 
 equally well if the muons are observed at the distance $L$ using any high 
 efficiency detector such as, for example, a scintillation counter. 
 In contrast, the rate of neutrino oscillation events is severely limited by
 the very small neutrino interaction cross section.

\SECTION{\bf{Neutrino oscillations following muon decay at rest or beta-decays
 in nuclear reactors}}
  The formula describing neutrino oscillations
 $\overline{\nu}_{\mu} \rightarrow \overline{\nu}_e$
 following the decay at rest of a $\mu^+$, $\mu^+ \rightarrow e^+ \nu_e \overline{\nu}_{\mu}$
 is easily derived from
 the similar formula where a $\nu_{\mu}$ is produced by $\pi^+$ decay at rest (2.25). Because the 
 neutrino momentum spectrum is continous, smearing effects due to the finite muon lifetime may
 be neglected from the outset. The phase increment associated with the neutrino path is then given 
 by Eqn(2.15) with the replacements $P_0 \rightarrow P_{\overline{\nu}}$ and $\delta_{\pi}, \delta_i
 \rightarrow 0$, where $ P_{\overline{\nu}}$ is the antineutrino momentum. The phase increment
 of the decaying muon is given by the same replacements in Eqn(2.16) and, in addition,
 $m_{\pi} \rightarrow m_{\mu}$ and $\Gamma_{\pi} \rightarrow \Gamma_{\mu}$.
 The formula, analogous to Eqn(2.26), for the time-averaged
 probability to detect the process $ \overline{\nu}_e p \rightarrow e^+ n$ at a distance L
 from the muon decay point is then:

\begin{eqnarray}
P(e^+n,\mu|L) & = & \frac{ |<e^+n|T|p \overline{\nu}_e>|^2
 |<\nu_e \overline{\nu}_{\mu} e^+|T|\mu^+>|^2}{\Gamma_{\mu}}
2 \sin^2\theta \cos^2\theta \nonumber \\
  &  & \times \left\{1-\exp[-\frac{\Gamma_{\mu} \Delta m^2}{4 P_{\overline{\nu}}^2}L]
\cos \left[\frac{\Delta m^2}{ P_{\overline{\nu}}}\left(
 \frac{m_{\mu}}{2 P_{\overline{\nu}}}-1\right)L \right] \right\}
\end{eqnarray} 
\par The standard neutrino oscillation formula, hitherto used in the analysis of all 
 experiments, has instead the expression $\Delta m^2 L/(2 P_{\overline{\nu}})$ for
 the argument of the cosine term in Eqn(3.1). Denoting my $\Delta m^2_S$ the value 
 of $\Delta m^2$ obtained using the standard formula, and $\Delta m^2_{FP}$ that
 obtained using the Feynman Path (FP) formula (3.1) then:
\begin{equation}
 \Delta m^2_{FP} = \frac{\Delta m^2_S}{\frac{m_{\mu}}{ P_{\overline{\nu}}}-2}
\end{equation}
 For a typical value of $P_{\overline{\nu}}$ of 45 MeV, Eqn(3.2) implies that
 $\Delta m^2_{FP} \simeq  2.9\Delta m^2_S$. Thus the $\overline{\nu}_{\mu}$ oscillation 
 signal from $\mu^+$ decays at rest reported by the LNSD Collaboration~\cite{LNSD}
 corresponding to $\Delta m^2_S \simeq 0.5$ (eV)$^2$ for $\sin^2 2 \theta \simeq 0.02$
 implies $\Delta m^2_{FP} \simeq 1.5$ (eV)$^2$ for a similar mixing angle.
\par For the case of $\beta$-decay:
\[ N(A,Z) \rightarrow  N(A,Z+1)e^-\overline{\nu}_e  \]
 $m_{\pi}$ in the first line of Eqn(2.16) is replaced by $E_{\beta}$, the total energy
 release in the $\beta$-decay process:
\begin{equation}
E_{\beta} = M_N(A,Z)-M_N(A,Z+1)
\end{equation}
where $M_N(A,Z)$ and $M_N(A,Z+1)$ are the masses of the parent and daughter nuclei.
That the phase advance of an unstable state, over a time, $\Delta t$, is given by
$\exp(-iE^* \Delta t)$ where $E^*$ is the excitation energy of the state, is
readily shown by the application of time-dependent perturbation theory
 to the Schr\"{o}dinger equation~\cite{Schiff}. A more intuitive derivation
 of this result 
 has been given in Ref.~\cite{LLB}. In the present
 case, $E^* = E_{\beta}$. Omitting the
lifetime damping correction, which is about eight orders of magnitude
 smaller than for pion decay, given a typical $\beta$-decay lifetime
 of a few seconds, the time-averaged probabilty to detect the $\overline{\nu}_e$
 via the process $\overline{\nu}_e p \rightarrow e^+n$, at distance $L$ from the
 decay point is given by the formula, derived in a similar way to Eqns(2.26) 
 and (3.1):
\begin{eqnarray}
P(e^+n,\beta|L) & = & \frac{ |<e^+n|T|p \overline{\nu}_e>|^2
 |<e^- \overline{\nu}_e N(A,Z+1)|T|N(A,Z)>|^2}{\Gamma_{\beta}} \nonumber \\
  &  & \times \left\{\sin^4 \theta + \cos^4 \theta + 2 \sin^2 \theta \cos^2 \theta
\cos \left[\frac{\Delta m^2}{ P_{\overline{\nu}}}\left(
 \frac{E_{\beta}}{2 P_{\overline{\nu}}}-1\right)L \right] \right\}
\end{eqnarray}
where $\Gamma_{\beta}^{-1} = \tau_{\beta}$ the lifetime of the unstable
 nucleus $N(A,Z)$. Until now, all experiments have used the standard 
 expression $\Delta m^2 L/(2 P_{\overline{\nu}})$ for the neutrino oscillation phase.
 The values of $\Delta m^2$ found should be scaled by the factor 
 $(E_{\beta}/P_{\overline{\nu}} -2)^{-1}$, suitably averaged over $P_{\overline{\nu}}$,
 to obtain the $\Delta m^2$ given by the Feynman Path formula (3.4).

\SECTION{\bf{Neutrino and muon oscillations following pion decay in flight}}

  In this Section, the decays in flight of a $\pi^+$ beam with energy $E_{\pi} \gg m_{\pi}$
 into $\mu^+ \nu_{\mu}$ are considered. As the analysis of the effects of the physical
 pion and muon masses have been shown above to give negligible corrections to the
 $L$ dependence of the oscillation formulae, for the case of decays at rest, such 
 effects will be neglected in this discussion of in-flight decays. The overall structure
 of the path amplitudes for neutrinos and muons is the same as for decays at rest (see Eqns(2.1)
 and (2.35)). However, for in-flight decays, in order to calculate the interfering paths
 originating at different and terminating at common space-time points, the two-dimensional
 spatial geometry of the problem must be properly taken into account. 
 \par In Fig.3 a pion decays at A into the 1 mass eigenstate, the neutrino being 
 emitted at an angle $\theta_1$ in the lab system relative to the pion flight direction.
 If $m_1 > m_2$ a later pion decay into the 2 mass eigenstate at the angle $\theta_1+\delta$
 may give a path such that both eigenstates arrive at the point B at the same time. A neutrino
 interaction $\nu_en \rightarrow e^- p$, or $\nu_{\mu}n \rightarrow e^- p$ occuring at this
 space-time point will than be sensitive to interference between amplitudes corresponding to the
 paths AB and ACB. The geometry of the triangle ABC and the condition that the 1 and 2 
 neutrino mass eigenstates arrive at B at the same time gives the following condition
 on their velocities:
\begin{equation}
\frac{v_1(\nu)}{v_2(\nu)} = \frac{\sin\theta_2}{\sin\theta_1}- \frac{v_1(\nu)}{v_{\pi}}
  \frac{\sin(\theta_2-\theta_1)}{\sin\theta_1} 
\end{equation}
Expanding to first order in the small quantity $\delta = \theta_2-\theta_1$, rearranging 
 and neglecting terms of $O(m_i^4)$ gives:
\begin{equation}
v_2(\nu)-v_1(\nu) = \frac{\Delta m^2}{2 E_{\nu}^2} = \frac{\delta}{\sin\theta_1}
\left[\frac{1-v_{\pi}\cos\theta_1}{v_{\pi}}\right]  
\end{equation}
where the relation:
\begin{equation}
 v_i(\nu) = 1 -\frac{m_i^2}{2 E_{\nu}^2} + O(m_i^4)  
\end{equation}
has been used.
Rearranging Eqn(4.2):
\begin{equation}
 \delta = \frac{\Delta m^2}{2 E_{\nu}^2}\left[\frac{v_{\pi}\sin\theta_1}
{1-v_{\pi}\cos\theta_1}\right]  
\end{equation} 
The difference in phase of the neutrino paths AB and CB is (see Eqn(2.15)):
\begin{equation}
 \phi^{\nu}_{12} = \frac{m_1^2 AB}{P_1} - \frac{m_2^2 CB}{P_2} + O(m_1^4,m_2^4)
\end{equation}
 Since the angle $\delta$ is $\simeq \Delta m^2$, the difference between AB and CD
 is of the same order, and so:  
\begin{equation}
 \phi^{\nu}_{12} = \frac{\Delta m^2 L}{\cos\theta_1 E_{\nu}} + O(m_1^4,m_2^4)
\end{equation} 
 where $ P_1 \simeq P_2 \simeq E_{\nu}$, the measured neutrino energy. From the
 geometry of the triangle ABC: 
 \begin{equation}
\frac{AC}{\sin \delta} = \frac{AB}{\sin\theta_2}= \frac{L}{\cos\theta_1 \sin(\theta_1+\delta)}
\end{equation}
 So, to first order in $\delta$, and using Eqn(4.4):
 \begin{equation}
 AC \equiv \Delta x_{\pi} = \frac{L \delta}{\cos\theta_1 \sin\theta_1} =
 \frac{\Delta m^2 L}{2 E_{\nu}^2 \cos\theta_1}\frac{v_{\pi}}{(1-v_{\pi}\cos\theta_1)}
\end{equation}
and    
 \begin{equation}
\Delta t_{\pi} =\frac{\Delta x_{\pi}}{v_{\pi}}=
 \frac{\Delta m^2 L}{2 E_{\nu}^2 \cos\theta_1}\frac{1}{(1-v_{\pi}\cos\theta_1)}
\end{equation}
Eqns(4.8) and (4.9) give, for the phase increment of the pion path:
 \begin{equation}
\Delta \phi^{\pi} = m_{\pi}(\tau_2-\tau_1) = m_{\pi} \Delta \tau = E_{\pi}\Delta t_{\pi}
-p_{\pi}\Delta x_{\pi} = \frac{\Delta m^2 E_{\pi} L}{2 E_{\nu}^2 \cos\theta_1}\frac{(1-v_{\pi}^2)}
 {(1-v_{\pi} \cos \theta_1)}
\end{equation}
 In Eqn(4.10), the Lorentz invariant character of the propagator phase is used. Setting
 $\cos \theta_1 = 1$ and $v_{\pi} = 0$, the pion phase increment of Eqn(2.25) is 
 recovered. Eqns(4.6) and (4.10) give, for the total phase difference of the paths 
 AB, ACB:
 \begin{equation}
\phi^{\nu,\pi}_{12} = \Delta \phi^{AB} - \Delta \phi^{ACB} =  \phi^{\nu}_{12}
 - \Delta \phi^{\pi} = \frac{\Delta m^2 L}{\cos\theta_1 E_{\nu}}
\left[1- \frac{E_{\pi}}{2 E_{\nu} }\frac{(1-v_{\pi}^2)}
 {(1-v_{\pi} \cos \theta_1)}\right]
\end{equation}
Using the expressions, valid in the ultra-relativistic (UR) limit where $v_{\pi} 
 \simeq 1$ :
 \begin{equation}
1-v_{\pi} \cos \theta_1 = \frac{m_{\pi}^2}{E_{\pi}^2}\frac{1}{(1+\cos\theta^{\ast}_{\nu})}
\end{equation} 
and
 \begin{equation} 
 E_{\nu} = \frac{E_{\pi}(m_{\pi}^2-m_{\mu}^2)}{2 m_{\pi}^2}
 {(1+\cos\theta^{\ast}_{\nu})}
\end{equation}
where $\theta^{\ast}_{\nu}$ is the angle between the directions of the pion and neutrino momentum
 vectors in the pion rest frame, Eqn(4.11) may be rewritten as:
 \begin{equation}
\phi^{\nu,\pi}_{12} = -\frac{\Delta m^2}{\cos\theta_1 E_{\nu}}\frac{m_{\mu}^2}
 {(m_{\pi}^2-m_{\mu}^2)} L
\end{equation} 
 For $\theta_1 = 0$ the oscillation phase is the same as for pion decay at rest 
 (see Eqn(2.26)) since in the latter case, $E_{\nu}\simeq P_{0} = (m_{\pi}^2-m_{\mu}^2)/(2 m_{\pi})$.
 Using Eqn(4.14), the probability to observe a neutrino interaction, at point B, produced by
 the decay product of a pion decay occuring within a region of length, $l_{Dec}$, ($\ll L$) centered at the 
 point A, in a beam of energy $E_{\pi}$, is given by a formula analagous to Eqn(2.26):
\begin{eqnarray}
P(e^-p|L,\theta_1) & = & \frac{l_{Dec} m_{\pi} \Gamma_{\pi}}{E_{\pi}} |<e^-p|T|n\nu_e>|^2
 |<\nu_{\mu} \mu^+|T|\pi^+>|^2
 \sin^2\theta \cos^2\theta \nonumber \\
  &  & \times \left\{1-
\cos\frac{ m_{\mu}^2 \Delta m^2 L}{( m_{\pi}^2- m_{\mu}^2) E_{\nu}\cos\theta_1} \right\} 
\end{eqnarray}
As in the case of pion decay at rest, Eqn(2.26), the oscillation phase differs by the
 factor $2 m_{\mu}^2/( m_{\pi}^2- m_{\mu}^2) = 2.685$
 from that given by the standard formula.
\par The derivation of the formula describing muon oscillations following pion decays
 in flight is very similar to that just given for neutrino oscillations. The condition
 on the velocities so that the muons recoiling against the different neutrino mass
 eigenstates arrive at the point B (see Fig.3) at the same time, is given by a formula
 analagous to (4.2):
\begin{equation}
\Delta v(\mu)= v_2(\mu)-v_1(\mu) = \frac{ v_1(\mu)[v_1(\mu)-v_{\pi}\cos\theta_1]}
{v_{\pi}\sin\theta_1}\delta  
\end{equation} 
 The formula relating the neutrino masses to the muon velocities is, however, more difficult
 to derive than the corresponding relation for neutrinos, (4.3), as the decay muons are not
 ultra-relativistic in the pion rest frame. The details of this calculation are given in 
 Appendix B. The result is:
\begin{equation}
\Delta v(\mu) = \frac{m_{\mu}^2 (m_{\pi}^2+m_{\mu}^2)\Delta m^2}{E_{\mu}^2(m_{\pi}^2-m_{\mu}^2)^2}
\left(1-\frac{2 m_{\mu}^2 E_{\pi}}{(m_{\pi}^2+m_{\mu}^2) E_{\mu}}\right)+O(m_1^4,m_2^4)  
\end{equation} 
Using Eqn(4.16) and the relation, valid to first order in $\delta$,:
\begin{equation}
\Delta t = \frac{ L \delta }{v_{\pi} \cos\theta_1 \sin\theta_1 } 
\end{equation}  
where $\Delta t$ is the flight time of the pion from A to C in Fig.3 (and
 also the difference in the times-of-flight of the muons recoiling against
 the two neutrino eigenstates), the angle $\delta$ may be eliminated 
 to yield:
\begin{equation}
 \Delta t = \frac{\Delta v(\mu) L}{v_1(\mu) \cos \theta_1[v_1(\mu)-v_{\pi}\cos\theta_1]}  
\end{equation} 
Using now the kinematical relation (see Appendix B):
\begin{equation}
v_1(\mu)-v_{\pi}\cos\theta_1 = \frac{(m_{\pi}^2+m_{\mu}^2)}{2E_{\pi}E_{\mu}}
\left(1-\frac{ 2 m_{\mu}^2 E_{\pi}}{(m_{\pi}^2+m_{\mu}^2) E_{\mu}}\right) 
\end{equation}
and the expression for the phase difference of the paths AB and ACB:
 \begin{equation}
\phi^{\mu,\pi}_{12} = \Delta \phi^{AB} - \Delta \phi^{ACB} = 
  \Delta t\left(\frac{m_{\mu}^2}{E_{\mu}}-  \frac{m_{\pi}^2}{E_{\pi}} \right)
\end{equation}
 together with Eqn(4.19), it is found, taking the UR limit, where  $v_1(\mu), v_{\pi} \simeq 1$, 
 that
 \begin{equation}
\phi^{\mu,\pi}_{12} = \frac{ 2 m_{\mu}^2 \Delta m^2}{E_{\mu}^2 (m_{\pi}^2-m_{\mu}^2)^2}
\left[\frac{(m_{\mu}^2 E_{\pi}- m_{\pi}^2 E_{\mu})L}{\cos\theta_1}\right]
\end{equation}
 The probability of detecting a muon decay at B is then:
\begin{eqnarray}
P(e^+\nu_e\overline{\nu}_{\mu}|L,\theta_1) & = & \frac{l_{Dec} m_{\pi} \Gamma_{\pi}}{E_{\pi}}
|<e^+\nu_e\overline{\nu}_{\mu}|T|\mu^+>|^2 \nonumber \\
  &  & 
 \times \exp \left[-\frac{\Gamma_{\mu} m_{\mu}}{E_{\mu}\cos\theta_1} L \right] |<\nu_{\mu} \mu^+|T|\pi^+>|^2
 \nonumber \\
  &  & \times \left\{1-\sin 2 \theta
 \cos    \frac{ 2 m_{\mu}^2 \Delta m^2}{E_{\mu}^2 (m_{\pi}^2-m_{\mu}^2)^2)}
\left[\frac{(m_{\mu}^2 E_{\pi}- m_{\pi}^2 E_{\mu})L}{\cos\theta_1}\right] \right\} 
\end{eqnarray}
where $l_{Dec}$ is defined in the same way as in Eqn(4.13).

\SECTION{\bf{Discussion}}
 The quantum mechanics of neutrino oscillations has been surveyed in two
 recent review articles~\cite{Zralek,GiuKi}, where further extensive
 lists of references may be found.
 \par The standard derivation of the neutrino oscillation phase will first be
 considered, following the treatment of Ref.~\cite{BiPet}, but using the 
 notation of the present paper. The calculation is performed in terms of the
 `flavour eigenstates of the neutrino', $| \alpha >$, which in the case of
 pion decay is, at $t=0$, essentially the pure state $ \alpha = \nu_{\mu}$
 corresponding to the superposition of the mass eigenstates $|\nu_1>$,
 $|\nu_2>$  given in Eqn(2.9) above. This flavour eigenstate is
 assumed to evolve with laboratoy time, $t$, according to fixed energy solutions of
 the non-relativistic
 Schr\"{o}dinger equation: 
 \begin{equation}
 | \alpha, t> = -\sin \theta e^{-iE_1t}| \nu_1, p> + \cos \theta e^{-iE_2t}| \nu_2, p>
\end{equation}
where  $|\nu_i, p>$ are mass eigenstates of fixed momentum $p$,
 and $E_1$, $E_2$ are the laboratory energies of the neutrino mass eigenstates.
 The amplitude for transition into the state $|\nu_e, p>$ at time $t$ is then, using
 Eqns(2.8), (2.9): 
 \begin{equation}
 <\nu_e, p|\alpha,t> = \sin \theta \cos \theta \left( - e^{-iE_1t} + e^{-iE_2t} \right)
 \end{equation}
 Assuming now that the neutrinos have the same momentum but different energies :
 \begin{equation}
  E_i = \sqrt{p^2 +m_i^2} = p + \frac{m_i^2}{2p} + O(m_i^4)
 \end{equation}
 and using (5.2) and (5.3), the probability of the flavour state $\nu_e$ at time
 $t$ is found to be:
 \begin{equation}
  P(\nu_e, t) = |<\nu_e, p|\alpha,t>|^2 = 2 \cos^2 \theta \sin^2 \theta
 \left(1 + \cos \left[\frac{(m_1^2- m_2^2)}{2 p}t \right] \right)
\end{equation}
 Finally, since the velocity difference of the neutrino mass eigenstates is
 O($\Delta m^2$), then, to the same order in the oscillation phase,
 the replacement $t \rightarrow L$ can be made in Eqn(5.4) to yield the standard
 oscillation phase of Eqn(1.1).

\par The following comments may be made on this derivation:
 \begin{itemize}
 \item[(i)]  The time evolution of the neutrino mass eigenstates in Eqn(5.1) according
   to the Schr\"{o}dinger equation yields a non-Lorentz-invariant
  phase $\simeq Et$, to be compared with the Lorentz-invariant phase
 $\simeq m^2t/E$ given in Eqn(2.12) above. Although the two expressions agree
 in the non-relativistic limit $E \simeq m$ it is clearly inappropriate to use
 this limit for the description of neutrino oscillation experiments. It may be
 noted that the Lorentz-invariant phase is robust relative to different 
 kinematical approximations. The same result is obtained to order $m^2$
 for the phase of spatial oscillations independent of whether the neutrinos
 are assumed to have equal momenta or energies. This is not 
 true in the non-relativistic limit. Assuming equal momenta gives the standard
 result of Eqn(1.1), whereas the equal energy hypothesis results in a 
 vanishing oscillation phase. 
   A contrast may be noted here with the standard 
   treatment of neutral kaon oscillations, which follows closely the derivation
  in Eqns(5.1) to (5.4) above, except that the particle phases are assumed to
  evolve with time according to $\exp [-im \tau]$ where $m$ is the particle
  mass and $\tau$ is its proper time, in agreement with Eqn(2.12).
   
 \item[(ii)] As pointed out in Ref.~\cite{Win}, the different neutrino
 mass eigenstates do not have equal momenta as assumed in Eqns(5.1) and (5.3).
  The approximation of assuming equal momenta might be
 justified if the fractional change in the momentum of the neutrino due to a
 non-vanishing mass were much less than that of the energy. However, in the 
 case of pion decay as is readily shown from Eqns(2.4) and (2.6) above,
 the ratio of the fractional shift in momentum to that in energy is actually
 $(m_{\pi}^2+m_{\mu}^2)/(m_{\pi}^2-m_{\mu}^2) = 3.67$; so, in fact, the
 opposite is the case.  
 \item[(iii)] To derive (5.2) from (5.1), it is necessary to introduce the
 concept of a `flavour momentum eigenstate' which cannot be defined in
  a theoretically consistent manner~\cite{GKL1}.
 \item[ (iv)] What are the physical meanings of $t$, $p$ in Eqn(5.1)?
   In Eqn(5.2), it is assumed that the neutrino mass eigenstates are both
   produced, and both detected, at the same times. Thus both have the same
   time-of-flight $t$.  
  The momentum $p$ cannot be the same for both eigenstates, as assumed
  in Eqn(5.3), if both energy
  and momentum are conserved in the decay process. For any given value
  of the laboratory time $t$ the different neutrino mass eigenstates
  must be at different space-time positions because they have different
  velocities\footnote{This is true not only in the case of energy-momentum
  conservation, but also if it is assumed, as in the derivation of the 
  standard formula, that the neutrinos have the same momentum but
  different energies.}, if it is assumed that both mass eigenstates are
  produced at the same time. It then follows that the $\nu_e$ part
  of the different mass eigenstates cannot be probed at, some space-time point,
  by a neutrino interaction, whereas the latter must clearly occur
  at some unique and specific point in every event. 
 \item[(v)] The historical development of the calculation of the neutrino
  oscillation phase is of some interest. The first published prediction
  ~\cite{GribPont} actually obtained a phase a factor two larger than
   Eqn(1.1) {\it i.e.} in agreement with the covariant calculation of
   Ref.\cite{Moh1}(before the introduction of `wave packets') and
   with the contribution from neutrino propagation found in the 
   present paper. This prediction was later used, for example,
   in Ref.\cite{BilPont}. The derivation sketched above, leading to
   the standard result of Eqn(1.1) was later given in Ref.\cite{FritMink}.
   A subsequent paper~\cite{BilPont1} by the authors of Ref.\cite{BilPont},
   published shortly afterwards, cited both Ref.\cite{GribPont} and
   Ref.\cite{FritMink}, but used now the prediction of the latter paper.
   No comment was made on the factor of two difference in the two
   calculations. In a later review article~\cite{BilPont2}
   by the authors of Ref.\cite{BilPont} a calculation similiar to
   that of Ref.\cite{FritMink} was presented in detail. Subsequently,
   all neutrino oscillation experiments have been analysed 
   on the assumption of the standard oscillation phase of Eqn(1.1).

\end{itemize}
   \par It may be thought that the kinematical
    and geometrical inconsistencies mentioned in points
    (ii) and (iv) above result from a too classical approach to the problem.
 After all,
    what does it mean, in quantum mechanics, to talk about the `position'
    and `velocity' of a particle, in view of the Heisenberg Uncertainty
    Relations~\cite{Heis1}? This point will become clear later in the
    present discussion, but first, following the original suggestion
    of Ref.~\cite{Kayser}, and, as done in almost all subsequent work
    on the quantum mechanics of neutrino oscillations, the `wave packet'
    description of the neutrino mass eigenstates will be considered.
    In this approach, both the `source' and also possibly the `detector'
    in the neutrino oscillation experiment are described by coherent
    spatial wave packets. Here the `source' wavepacket treatment in
    the covariant approach of Ref.~\cite{Moh1} will be briefly
     sketched. After discussing the results obtained, and comparing them
    with those of the present paper, the general consistency of the
    wave packet approach with the fundamental quantum mechanical formula (1.2)
    will be examined.   
    \par The basic idea of the wave packet approach of Ref.~\cite{Moh1}
     is to replace the neutrino propagator (2.11) in the path amplitude
     by a four-dimensional convolution of the propagator with a 
     `source wavefunction' which, presumably, describes the space-time
     position of the decaying pion. For mathematical convenience this
     wavefunction is taken to have a Gaussian form with spatial
     and temporal widths $\sigma_x$ and $\sigma_t$ respectively.
    Thus, the neutrino propagator $D$ is replaced by $\tilde{D}$ where:
\begin{equation}
  \tilde{D}(x_f-x_i,m_j) = \int d^4x D(x_f-x,,m_j) \psi_{in}(x -x_i,j)
\end{equation}
 where $x_i$ and $x_f$ are 4-vectors and
\begin{equation}
  \psi_{in}(x -x_i,j) =  N_0 \exp\left[-ip_j \cdot (x-x_i)-\frac{(\vec{x}-\vec{x}_i)^2}
   {4 \sigma_x^2}- \frac{(t-t_i)^2}{4 \sigma_t^2} \right]
\end{equation}
 The integral in (5.5) was performed by the stationary phase method, yielding
  the result (up to multiplicative and particle flux factors), and here assuming,
  for simplicity, one dimensional spatial geometry:
\begin{equation}
  \tilde{D}(\Delta x,\Delta t, m_j) = \exp\left[ -i\left(\frac{m_j^2}{E_j}\Delta t-P_j(\Delta x- v_j 
  \Delta t)\right)-\frac{(\Delta x- v_j \Delta t)^2}{4(\sigma_x^2+v_j^2 \sigma_j^2)}\right]
\end{equation}
 For the case of $\nu_{\mu} \rightarrow \nu_e$ oscillations, following $\pi^+$ decay at rest, the
 probability to observe flavour $\nu_e$ at time $t$ and distance $x$ is given by:
\begin{equation}
P(\nu_e,\Delta x , \Delta t) = \cos^2\theta \sin^2\theta \left|-\tilde{D}(\Delta x,\Delta t, m_1)
 +\tilde{D}(\Delta x,\Delta t, m_2) \right|^2   
\end{equation}
 Performing the integral over $\Delta t$ and making the ultra-relativistic approximation
  $v_1, v_2 \simeq 1$ yields finally, with $\Delta x = L$:

\begin{eqnarray}
 P(\nu_e, L) & = & 2 \cos^2\theta \sin^2\theta \left(1+    \right. \nonumber \\ 
     &  & \exp\left[-\frac{\Delta m^4 (\sigma_x^2 + \sigma_t^2)}{2 m_{\pi}^2}
      - \frac{\Delta m^4
     (m_{\pi}^2+m_{\mu}^2)^2 L^2}{4(m_{\pi}^2-m_{\mu}^2)^4 (\sigma_x^2 + \sigma_t^2)}
  \right] \nonumber \\ 
  &  & \times  \cos \frac{\Delta m^2 L}{2 P_0} )
\end{eqnarray}
 It can be seen that the oscillation phase in Eqn(5.9) is the same
 as standard one of Eqn(1.1).
 The exponential damping factors in Eqn(5.9) are the same as those originally found
 in Ref.~\cite{GKL}
 for spatial wave packets ($\sigma_t = 0$).
 Considering now only spatial wave packets and using the property $\sigma_x \sigma_p = 1/2$ 
 derived from the Fourier transform of a Gaussian, the two terms in the exponential 
 damping factor may be written as:
\begin{equation}
 F_p =\exp \left[ -\frac{\Delta m^4 }{ 8 m_{\pi}^2 \sigma_p^2 } \right ]
\end{equation}
and    
\begin{equation}
 F_x =\exp \left[ -\frac{\Delta m^4 (m_{\pi}^2+m_{\mu}^2)^2 L^2 }
      {4 (m_{\pi}^2-m_{\mu}^2)^4 \sigma_x^2}\right]
\end{equation} 
 \par The spatial damping factor $F_x$ is usually interpreted in terms of a 
 `coherence length'`\cite{Nussinov}. If 
 \begin{equation}
   L \gg \frac{2 (m_{\pi}^2-m_{\mu}^2)^2 \sigma_x}
     { \Delta m^2 (m_{\pi}^2+m_{\mu}^2)}
 \end{equation}
 then $F_x \ll 1$ and the neutrino oscillation term is strongly suppressed. Eqn(5.12)
 expresses the condition that oscillations are only observed provided that the
 wave packets overlap. Since $\Delta v \simeq \Delta m^2$ the separation of
  the wave packets is $ \simeq \Delta v L \simeq \Delta m^2 L$, so that Eqn(5.12)
 is equivalent to $ \Delta v  L \gg  \sigma_x$ (no wave packet overlap).
  \par The damping
  factor $F_p$ is typically interpreted~\cite{GKL2} in terms of the `Heisenberg
  Uncertainty Principle'. This factor is small, unless the difference in mass
 of the eigenstates is much less than $2^{\frac{5}{4}}\sqrt{m_{\pi} \sigma_p}$, so
 it is argued that only for wide momentum wave packets can neutrino oscillations
 be observed, whereas in the contrary case, when the mass eigenstates are distinguishable,
 the interference effect vanishes. In the case of pion decay the difference in 
 momentum of the two interfering mass eigenstates comes only from
 the $\delta_i$ term in Eqn(2.4), as $\delta_{\pi}$, being a property of 
 the common initial state, is the same for both eigenstates. The neutrino
 momentum smearing in pion decay is then estimated from Eqn(2.4) as:
 \[ \sigma_p = |P_i-P_0| = \frac{2 P_0 \delta_i m_{\mu}}{m_{\pi}^2-m_{\mu}^2} 
   \simeq \frac{ \Gamma_{\mu} m_{\mu}}{ 2 m_{\pi}} = 1.14 \times 10^{-10} \rm{eV} \]
  For $\Delta m^2 = 1(\rm{eV})^2$ the value of $F_p$ is found to be $\exp(-500)
  \simeq 7.1 \times 10^{-218}$ giving complete suppression of neutrino oscillations.
  This prediction is in clear contradiction with the tiny damping corrections
  found in the path amplitude analysis in of Section 2 above. 
  \par The physical
  relevance of the damping factors $F_x$ and $F_p$ will be reviewed after
  the examination of the physical meaning of the wave packet convolution
  formula (5.5) which now follows.
  \par The convolution formula (5.5) may be re-written so as to 
    isolate the contribution from the region:
    \[ x_1'-\frac{\Delta x_1'}{2} < x_1 <  x_1'+\frac{\Delta x_1'}{2}  \]
 where $x_1'$ is an arbitary fixed value of the coordinate $x_1$ (say, the spatial
    position of source pion parallel to the direction of neutrino propagation):
\newpage
\begin{eqnarray}
\tilde{D}(x_f-x_i,m_1) & = & \int D\psi_{in}\left[\int_{-\infty}^
 {x_1'-\frac{\Delta x_1'}{2}} dx_1+\int^{\infty}_{x_1'+\frac{\Delta x_1'}{2}}dx_1
 \right]d^3x + \Delta x_1'\int \left[D\psi_{in}\right]_{x_1=x_1'}d^3x \nonumber \\
  & \equiv &  \int_{(-\Delta x_1')} D\psi_{in}d^4x+
  \Delta x_1'\overline{D\psi_{in}}(x_1')
\end{eqnarray}
 Here the $x_2$, $x_3$, $x_4$ averaged quantity $\overline{D\psi_{in}}(x_1')$
 is a function of $(x_f)_k-(x_i)_k$, $k=2,3,4$ as well as $(x_f)_1-x_1'$  and
 $x_1'-(x_i)_1$. Similarly:
 \begin{equation}
\tilde{D}(x_f-x_i,m_2) = \int_{(-\Delta x_1'')} D\psi_{in}d^4x+
  \Delta x_1''\overline{D\psi_{in}}(x_1'')
\end{equation}
 Now in calculating the oscillation probability (5.8), Eqns(5.13) and (5.14)
  will give a contribution to the interference term from the terms proportional
  to $\Delta x_1'$, $\Delta x_1''$ of:
 \begin{equation}
  I(\Delta x_1',\Delta x_1'') = \Delta x_1'\Delta x_1''2 {\it Re}
  \overline{D\psi_{in}}(x_1')^* \overline{D\psi_{in}}(x_1'')    
\end{equation}
 This contribution corresponds to interference between a pion decaying at an
 arbitary point $x_1=x_1'$ with another decaying at another arbitary point
 $x_1=x_1''$. This is in contradiction with Eqn(1.2), according to which,
 {\it the initial quantum state must be the same for all interfering 
  path amplitudes}. Referring to Fig 1., it can be seen that the
  different path amplitudes correspond to two (classically alternative)
 decay histories of the {\it same pion} which must clearly be at some
  unique spatial position at time $t_0$. `Each pion interferes only with
  itself. Interference between two {\it different} pions never
  occurs'~\cite{Dirac1}. Thus the convolution, at amplitude level, of
  the neutrino propagator with `source' and/or `detector' wavefunctions
  does not respect Heisenberg and Feynman's fundamental law of quantum mechanics
  (1.2). 
   \par The preceding discussion of the derivation of the standard formula 
    for the oscillation phase (1.1) in terms of `flavour eigenstates' revealed
      contradictions and inconsistencies
     if the neutrinos are assumed to follow classical space-time trajectories.
     The `source wave packet' treatment gives the standard result for the
     oscillation phase and predicts that the interference term will be
     more or less damped depending on the widths in configuration and 
     momentum space of the wave packets. However the convolution
     formula (5.5) is evidently at variance with Eqn(1.2) and the requirement
     that particles can interfere only `with themselves'. So do wave packets
     actually play a role in the correct quantum mechanical description
     of neutrino oscillations, as suggested in Ref.\cite{Kayser}? Are the
     packets actually constrained by the Heisenberg Uncertainty Relations?
     How do the properties of the wave packet effect the possiblity to 
     observe neutrino oscillations? The answers to all these questions 
     are contained in the results of the calculations presented in
     Section 2 above. They are now reviewed, with special emphasis
     on the basic assumptions made and the physical interpretation
     of the equations.     
       \par Referring again to Fig.1, in a) a single pion comes
     to rest in the stopping target T. The time of its passage is
     recorded by the counter C$_{\rm A}$, which thus defines the initial
     state as a $\pi^+$ at rest at time $t_0$. This pion, being
     an unstable particle, has a mass $W_{\pi}$ that is, in
      general, different from its most likely value which is 
     the pole mass $m_{\pi}$. What are shown in Fig.1b) and Fig.1c)
     are two different classical histories of this {\it very same pion}.
     In b) it decays into the neutrino mass eigenstate $\nu_1$ at time $t_1$,
     and in c) it decays into the neutrino mass eigenstate $\nu_2$ at
     time $t_2$. Because these are independent classical histories, the
     physical masses  $W_{\mu(1)}$, $W_{\mu(2)}$ of the muons recoiling
     against the mass eigenstates $\nu_1$ and  $\nu_2$ respectively, are,
     in general, not equal. Taking into account now exact energy-momentum
     conservation in the decay processes (appropriate because of the
     covariant formulation used throughout) the eigenstates $\nu_1$ and
     $\nu_2$ will have momenta which depend on $W_{\pi}$ and $W_{\mu(1)}$
     and $W_{\pi}$ and $W_{\mu(2)}$ respectively. These momenta are calculated
     in Eqn(2.4). The distributions of $W_{\pi}$ and $W_{\mu(i)}$ are
     determined by Breit-Wigner amplitudes, that are the Fourier transforms
     of the corresponding exponential decay laws, in terms of the decay
     widths $\Gamma_{\pi}$ and $\Gamma_{\mu}$ respectively.
     In accordance with Eqn(1.2),
     only the Breit-Wigner amplitudes corresponding to the physical muon
     masses are (coherently) integrated over at the amplitude level. The
     integral over $W_{\pi}$ (a property of the initial state) is 
     performed (incoherently) at the level of the transition probability.
     Because of the long lifetimes of the $\pi$ and $\mu$, the
      corresponding momentum wave packets are very narrow, so that all
     corrections resulting from integration over the resulting momentum
     spectra are found to give vanishingly small corrections.
\begin{table}
\begin{center}
\begin{tabular}{|c||c|c|c|c|} \hline 
  Source  &  $\Delta m = 1 {\rm eV}$ & $W_{\mu}$ &   $W_{\pi}$ & Doppler Effect \\
\hline
\hline
$\Delta P_{\nu}/P_{\nu}$ & 4.4$\times 10^{-16}$ & 3.8$\times 10^{-18}$
 & 3.3$\times 10^{-16}$ & 1.9$\times 10^{-5}$  \\ 
\hline
\end{tabular}
\caption[]{{\it Different contributions to neutrino momentum
   smearing in pion decay at rest (see text).
}}
\end{center}
\end{table}
     Indeed, as
     is evident from the discussion of Eqn(5.10) above, the width of the 
     momentum wave packet is much smaller than the difference in the 
      momenta of the eigenstates expected for experimentally interesting
     values of the neutrino mass difference (say,  $\Delta m^2 = 1 ({\rm eV})^2$). 
     However, contrary to the prediction of Eqn(5.10), this does not at all
     prevent the observation of neutrino oscillations. This is because
     the oscillations result from interference between amplitudes 
     corresponding to different propagation times, not momenta, of the
     neutrino mass eigenstates. Table 1 shows contributions
     to the fractional smearing of the neutrino momentum from
     different sources:
     \begin{itemize}
     \item[(1)] neutrino mass difference $\Delta m^2 = 1 ({\rm eV})^2$ (Eqn(2.4))  
     \item[(2)] coherent effect due to the physical muon mass
                ($\delta_i=\Gamma_{\mu}/2$ in Eqn(2.4))
     \item[(3)] incoherent effect due to the physical pion mass
                ($\delta_{\pi}=\Gamma_{\pi}/2$ in Eqn(2.4))
      \item[(4)] incoherent Doppler effect assuming $\overline{p}_{\pi}
                 = 2.6 \times 10^{-3} {\rm MeV}$
     \end{itemize}
      The pion mass effect is of the same order of magnitude as the 
      neutrino mass shift. The muon mass effect is two orders of
      magnitude smaller, while the Doppler effect at room temperature
      gives a shift eleven orders of magnitude larger than a (1eV)$^2$ 
      neutrino mass difference squared.

 \par According to the usual interpretation of the Heisenberg
  Uncertainty Principle, the neutrinos from pion decay, which, as has 
  been shown above, correspond to very narrow momentum wave packets,
  would be expected to have a very large spatial uncertainty. Indeed,
  interpreting the width of the coherent momentum wave packet 
  generated by the spread in $W_{\mu}$ according to the the momentum-space
  Uncertainty Relation $\Delta p \Delta x = 1$ gives 
  $\Delta x = 1.27$ km. Does this accurately represent the knowledge
  of the position of a decay neutrino obtainable in the experiment
  shown in Fig.1?  Without any experimental difficulty, the decay time
  of the pion can be measured with a precision of $10^{-10}$sec, 
  by detecting the decay muon. Thus, at any later time the
  distance\footnote{ Here, `distance' is defined without specifying
  the flight direction of the neutrino. Precise, simultaneous, 
  measurement of the flight direction of the recoil muon also
  determines, with a similar precision, the neutrino direction. Evidently
  the thickness of the stopping target may be chosen sufficiently
  small that its contribution to the neutrino position uncertainty
  is negligible.}  
  of the neutrino from the decay point is known
  with a precision of $c\times 10^{-10}\rm{cm} = 3$cm. This is a factor
  4$\times 10^4$ more precise than the `uncertainty' given by
  the Heisenberg relation. It is clear that the experimental 
  knowledge obtainable on the position of the neutrino is essentially
  classical, in agreement with the theoretical description in
  terms of classical particle trajectories in space-time. In this
  case the momentum-position Uncertainty Relation evidently
  does not reflect the possible experimental knowledge of the 
  position and momentum of the neutrino. This is because it
  does not take into account the prior knowledge that the
  mass of the neutrino is much less than that of the
  pion or muon, so that its velocity is, with 
  negligible uncertainty, c. In spite of this,
  a Heisenberg Uncertainty Relation is indeed respected in the 
  pion decay process. The Breit-Wigner amplitude that
  determines the coherent spread of neutrino momentum is
  just the Fourier transform of the exponential decay law of the 
  muon. The width parameter of the Breit-Wigner amplitude and the
  muon mean lifetime do indeed respect the {\it energy-time} Uncertainty Relation
  $\Gamma_{\mu} \tau_{\mu} =1$.
  \par It is then clear, from
  this careful analysis of neutrino oscillations following
  pion decay at rest, that, in contradiction to what
  has been almost universally assumed until now, 
  {\it the neutrinos are not described by a coherent spatial
   wave packet}. There is a coherent momentum wave packet,
   but it is only a kinematical consequence of a Breit Wigner
   amplitude. If for mathematical convenience, the momentum
   wave packet is represented by a Gaussian, a conjugate
    (and spurious) Gaussian spatial wave packet will be generated
   by Fourier transformation. Indeed, in the majority
  of wave packet treatments that have appeared in the literature,
   Gaussian momentum and spatial wave packets related by
   a Fourier transform with widths satisfying the `Uncertainty
   Relation' $\sigma_p \sigma_x =1/2$ have been used. 
   The above discussion shows that such a treatment does
   not correspond to the actual knowledge of the space-time events
   that constitute realistic neutrino oscillation experiments
    \footnote{It follows from this that discussions of quantum
    mechanical coherence in neutrino oscillations based on the properties
      of Gaussian wavepackets, as in Ref.~\cite{KNW}, do not address the
      actual physical basis of neutrino oscillation experiments. 
    A similar remark applies to the more recent and very extensive
    study of Ref.~\cite{Beuthe}. }. 
    \par The limitation on the detection distance $L$, for observation
   of neutrino oscillations, given by the damping factor $F^{\nu}(\Gamma_{\pi})$ of 
  Eqn(2.30) is easily understood in terms of the classical particle trajectories
  shown in Fig.1. For a given velocity difference, the time difference
  $t_2-t_1$ becomes very large when both neutrinos, in the
  alternative classical histories, are required to
  arrive simultaneously at a far distant detector. Because of the finite
   pion lifetime, however, the amplitude for pion decay at time $t_2$ is
   smaller than that at $t_1$ by the factor $\exp[-(t_2-t_1)\Gamma_{\pi}/2]$.
   Integrating over all decay times results in the exponential damping
   factor $F^{\nu}(\Gamma_{\pi})$ of Eqn(2.30). It is clear that, contrary
   to the damping factor $F_x$ of Eqn(5.11), the physical origin of the 
   $L$ dependent damping factor is quite unrelated to `wavepacket overlap'.
   \par For a given value of $t_2-t_1$ the coherent neutrino momentum
   spread originating in the Breit Wigner amplitude for $W_{\mu}$
   produces a corresponding velocity smearing that reduces the number
   of possible classical trajectories arriving at the detection event.
   This effect is taken into account in the integral shown in Eqn(2.23).
   The effect is shown, in Section 2 above, to be much smaller than the (already tiny)
   pion lifetime damping described above, and it is neglected in Eqn(2.26).
    \par Finally, as descibed after Eqn(2.25), the (incoherent) integration
   over the Breit Wigner function containing $W_{\pi}$ gives an additional
   damping correction to the interference term that is also very tiny compared
   to that due to the pion lifetime.
   \par As promised above, a final remark is now made on the physical interpretation
    of the damping
    factors (5.10) and (5.11) that have often been derived and discussed in
    the literature. Although as pointed out above, and to be further discussed
    below, the derivation of these factors is incorrect, as it results from
    coherent integration over different spatial positions of the neutrino
    source, similar factors do appear when performing the integrations
   over the (unobserved) neutrino momentum distributions. $F_x$ is replaced
   by $F^{\nu}(\Gamma_{\pi})$ and $F_p$  by the factor resulting from the
   coherent integration over the physical muon mass $W_{\mu}$. The 
   reason for the huge suppression factor predicted by Eqn(5.10), and the
    tiny one found in the path amplitude calculation, is that, in
    deriving (5.10) and (5.11), it is assumed that the neutrino
    eigenstates are both produced and detected at equal times. This will
    only be possible if both the hypothetical
    `wave packet overlap' is appreciable (Eqn(5.11)) and also the momentum
     smearing is sufficiently large that the time-of-flight differences
     due to the different neutrino masses are washed out (Eqn(5.10).
     In the path amplitude calculation interference and hence oscillations
     are made possible by different {\it decay times} of the source pion,
     and the damping factors analagous to $F_x$ and $F_p$ turn out
      to give vanishingly small corrections to the oscillation term.
\par It may be remarked that the physical interpretation of `neutrino oscillations'
  provided by the path amplitude description is different from the conventional
  one in terms of `flavour eigenstates'. In the latter the amplitudes of different
  flavours in the neutrino are supposed to vary harmonically as a function of time.
  In the amplitudes for the different physical processes in the path amplitude
  treatment there is, instead, no variation of lepton flavour in the propagating
  neutrinos. If the mass eigenstates are represented as superpositions of 
  flavour eigenstates using the inverses of Eqns(2.8) and (2.9), there is
   evidently no temporal variation of the lepton flavour composition. 
   Only in the detection process itself are the flavour eigenstates 
   projected out, and the interference effect occurs that is 
   described as `neutrino oscillations'. In the case of the 
   observation of the recoil muons no such projection takes place,
   but exactly similar interference effects are predicted to occur.
   As previously emphasised~\cite{SWS}, the `flavour oscillations'
   of neutrinos, neutral kaons and b-mesons are just special examples
   of the universal phenomenon of quantum mechanical superposition
   that is the basis of Eqn(1.2). 
      \par A more detailed critical review will now be made of previous
     treatments of the quantum mechanics of both neutrino and muon oscillations
     in the literature.
     \par By far the most widespread difference from the path amplitude treatment
     of the present paper is the non-respect of the basic quantum mechanical
     formula, (1.2), by the introduction of wave packets to describe `source'
     and/or `detector' particles[7,8,10,11,30-46]. Since in any
     practical neutrino oscillation experiment a single initial or final
      quantum state, as specified by Eqn(1.2), is not defined, but
     rather sets of initial and final states $I = \sum_l i_l$ and 
     $F = \sum_m f_m$ determined by experimental conditions, Eqn(1.2)
     may be generalised to:
  \begin{equation}
 P_{FI} = \sum_m \sum_l \left|\sum_{k_1} \sum_{k_2}...\sum_{k_n}\langle f_m| k_1 \rangle
\langle k_1| k_2 \rangle...\langle k_n| i_l \rangle \right|^2  
 \end{equation}
   to be contrasted with the formula used in the references cited above:
  \begin{equation}
 P_{fi} = \left|\sum_m \sum_l\ \sum_{k_1} \sum_{k_2}...\sum_{k_n}\langle \psi_f|f_m\rangle
 \langle f_m| k_1 \rangle
\langle k_1| k_2 \rangle...\langle k_n| i_l \rangle\langle i_l  |\psi_i\rangle \right|^2  
 \end{equation}
    Here, $\psi_i$ and $\psi_f$ are `source' and `detector' wave packets respectively.
   In Eqn(5.17) the initial states $i_l$ and final states $f_m$ of Eqn(5.16), that
   correspond to different spatial positions, and also, possibly, different
   kinematic properties,
   of the source particle or detection event, are convoluted with spatial and/or temporal 
 `wave packets', that, as 
   discussed above, are devoid of physical significance.
   Use of Eqn(5.17) leads to the absurd prediction (see, for example, Eqn(5.15) above)
   that {\it different} initial or final state particles interfere with each other.
   A possible reason for the widespread use of Eqn(5.17) instead of (5.16) may be understood
  following a remark of the author of Ref.~\cite{Campagne} concerning a `paradox' of the 
   complete quantum field theory calculation that takes into account, by a single
   invariant amplitude, production, propagation and detection of the neutrinos.
   It was noticed that, if the amplitude for the complete chain of processes is
   considered to correspond to one big Feynman diagram, then integration
   over the space-time coordinates of the initial and final states will
   reduce the exponential factors, containing the essential information on the
   interference phase, to energy-momentum conserving delta functions, and
   so no oscillations will be possible. A related remark was made by the
   authors of Ref.~\cite{DMOS} who stated that, as they were assuming 
   exact energy-momentum conservation, the integration over the space-time
   coordinates could be omitted. They still, however,
   (quite inconsistently, in view of the previous remark) retained the
   exponential factors containing the interference phase
   information. These considerations indicate a general confusion between
   momentum space Feynman diagram calculations, where it is indeed
  legitimate to integrate, at the amplitude level, over the
  unobserved space-time positions of the initial and final state particles,
   and the case of neutrino oscillations, where the amplitude
   must be defined in configuration space. In the latter case, it is
   the unobserved momenta of the propagating particles that should be
    integrated over, as is done in
   Eqns(2.1) and (2.35) above, and not the space-time positions of the
   `source' or `detector' particles, as in Eqn(5.17). 
   \par It is clear that exact energy-momentum conservation plays a
   crucial role in the path amplitude calculation. This is valid
   only in a fully covariant theory. Still, several authors, in spite
   of the ultra-relativistic nature of neutrinos, used a non-relativistic
   theory to describe the production, propagation and detection of
   neutrinos~\cite{Rich,KW,GMS,GMS1}. As is well known, in such
   `Old Fashioned Perturbation Theory'~\cite{HM} energy is not conserved
   at the level of propagators and so no precise analysis of the
   kinematics and the space-time configurations of the production and
   detection events, essential in the covariant path amplitude analysis, is
   possible.
   \par Even when, in some cases, the complete production, propagation
    and detection process
   of the neutrinos were described[35-37,39,40], equal neutrino energies
   [37,39,40],  equal neutrino momenta [36] or
    either~\cite{Beuthe}
   were assumed, in contradiction with energy-momentum conservation
   and a consistent space-time description of the production and detection 
   events. As follows directly from Eqn(5.3) (or the similar formula,
    for the neutrino momentum, obtained by assuming equal neutrino energies),
   in all the 
   above cited references, the standard formula (1.1) for the oscillation
   phase was obtained.
   \par An interesting discussion of the interplay between different
    kinematical assumptions (not respecting energy-momentum conservation)
    and the space-time description of the production and detection
   events was provided in Ref.~\cite{LDR}. This treatment was based on the
   Lorentz-invariant propagator phase of Eqn(2.12).
   By assuming either equal momentum
   or equal energy for the propagating neutrinos, but allowing {\it different
   times of propagation} for the two mass eigenstates, values of $\phi_{12}^{\nu}$
   agreeing with Eqns(2.15) and (2.25) above were found, {\it i.e.} differing by a factor
   two from the standard formula. Alteratively, assuming equal velocities,
   (and hence equal propagation times) the standard result (1.1) was obtained.
    In this latter case, however, the masses, momenta and energies of the
    neutrinos must be related, up to corrections of O($m_i^2$), according to:
     \begin{equation}
      \frac{m_1}{m_2} =   \frac{P_1}{P_2} = \frac{E_1}{E_2}
     \end{equation}
      Since the ratio of the neutrino masses may take, in general, any value,
     so must
    then the ratio of their momenta. For the case of neutrino production
    from pion decay at rest with $m_1,m_2 \ll m_{\mu}, m_{\pi}$ the 
     relation (5.18) is clearly incompatible with Eqn(2.4) which gives: 
     \begin{equation} 
       \frac{P_1}{P_2} = 1-\frac{\Delta m^2 (m_{\pi}^2+m_{\mu}^2)}
        {(m_{\pi}^2-m_{\mu}^2)^2}+ O(m_1^4,m_2^4)
     \end{equation}
     On the other hand, Eqn(5.19) is clearly compatible (up to corrections
     of  O($\Delta m^2$)) with the `equal momentum' hypothesis. There
     is similar compatiblity with the `equal energy' hypothesis. Even so,
     the authors of Ref.~\cite{LDR} recommended the use of the equal velocity
     hypothesis.  With the hindsight provided by the path amplitude
      analysis, in which the two neutrino mass eigenstates do indeed have 
     different propagation times, it can be seen that the kinematically
     consistent `equal momentum' and `equal energy' choices are good 
     approximations and the neutrino oscillation phase, resulting
      from the propagation of the neutrinos alone, is indeed a
      factor two larger than the prediction of the standard formula.

      \par Only the author of Ref.~\cite{Campagne} included the propagator of
     the decaying pion in the complete production-propagation-detection amplitude;
     compare Eqn(8) of Ref.~\cite{Campagne} with Eqn(2.1) above. However, no detailed
     space-time analysis of production and detection events was performed. Square
     spatial wave packets  for the `source' and `detector' were convoluted
     at amplitude level as in Eqn(5.17). As the neutrinos were assumed to have equal
     energies the standard result for the oscillation phase was obtained.

      \par Although, following Ref.~\cite{Rich}, most recent studies of the
     quantum mechanics of neutrino oscillations have considered the complete
     production-propagation-detection process, some authors still use,
     in spite of the criticisms of Ref.~\cite{GKL1}, the `flavour eigenstate' 
     description~\cite{GroLip,BV,BHV}. In the last two of these references a
     `quantum field theory' approach ia adopted, leading to an `exact' oscillation formula
     ~\cite{BHV} that does not make use of the usual ultra-relativistic
      approximation. This work did not include either exact kinematics, or an
      analysis of the space-time structure of production and detection events.
      Also the effect of the propagator of the source particle was not 
      considered. In Ref.~\cite{GroLip}
      the equal energy hypothesis was used and in Ref.~\cite{BV,BHV} the equal 
      momentum hypothesis. As a consequence, in all three cases the standard
      result was found for the oscillation phase in the ultra-relativistic 
      limit.   
       \par Correlated production and detection of neutrinos and muons produced
       in pion decay were considered in Ref.~\cite{SWS}. The introduction to this
       paper contains a valuable discussion of the universality
        of the `particle oscillation' phenomenon. It is pointed out that this
       is a consequence of the general principle of amplitude superposition in
        quantum mechanics, and so is not a special property of the 
        $\rm{K}^0-\overline{\rm{K}}^0$, $\rm{B}^0-\overline{\rm{B}}^0$
        and neutrino systems which are usually discussed in this context. 
        This paper used a covariant formalism that employed the `energy representation'
        of the space-time propagator. In the introduction, the important difference
        between Eqns(5.16) and (5.17) was also touched upon:
        \par` The reader will agree that one should not integrate over space
               if one is interested in spatial interference (or oscillation).'  
        \par Even so, in the amplitude for the correlated detection of the muon and
         neutrino (Eqn(2.10) of Ref.~\cite{SWS}) not only are the space-time positions
         of the production points of the neutrino and muon integrated over, but they are
         assumed to be at {\it different} space-time points. The propagator of the
         decaying pion is not included in the amplitude, and although exact 
         energy-momentum conservation is imposed, no space-time analysis
         of the production and decay points is performed. Correlated spatial
         oscillations of neutrinos and muons are predicted, though with interference
         phases different from the results of both the present paper and the 
         standard formula. Integrating over the spatial position of either
         the detected neutrino or muon destroys the interference term, so
         it is predicted that only correlated neutrino-muon oscillations may
         be observed, following pion decay, in contradiction with the results
         of the present paper, and, (in the case of neutrino oscillations)
         with all previous studies\footnote{Although this conclusion follows from Eqn(3.22) of
         Ref.~\cite{SWS}, it was not pointed out in the paper.}. Pion and muon lifetime
          effects were
         mentioned in Ref.~\cite{SWS}, but neither the role of the pion
         lifetime in enabling different propagation times for the neutrinos
         nor the momentum smearing, induced by the Fourier-transform-related
         Breit Wigner amplitudes, were discussed.
        \par The claim of Ref.~\cite{SWS})
         that correlated neutrino-muon oscillations should be observable in pion
         decay was questioned in Ref.~\cite{DMOS}. The authors of the latter paper
         attempted to draw conclusions on the possibility, or otherwise,
         of particle oscillations by using `plane waves', {\it i.e.} energy-momentum eigenfunctions.
         As is well known, such wavefunctions are not square integrable, and so can yield
         no spatial information. The probability to find a particle described by such
         a wave function in any finite spatial volume is zero. Due to the omission
          of the (infinite) normalisation constants of the wavefunctions many of the
          equations in Ref.~\cite{DMOS} are, as previously pointed out~\cite{SW},
          dimensionally incorrect. Momentum wavepackets for the decaying pion
          convoluted at amplitude level as in Eqn(5.17) were also discussed in 
           Ref.~\cite{DMOS}. Although exact energy-momentum conservation constraints 
           were used, it was assumed, as in Ref.~\cite{SWS},  that the muons and the
           different neutrino mass eigenstates are both produced and detected 
           at common points
            (Eqn(35) of Ref.~\cite{DMOS}). The latter assumption implies equal
            velocities, yielding the standard neutrino oscillation phase as
           well as the inconsistent kinematical relation (5.18). The authors of
           Ref.~\cite{DMOS} concluded that:
           \begin{itemize}
           \item[(a)] correlated $\mu-\nu$ oscillations of the type discussed in 
                      Ref.~\cite{SWS} could be observed, though with different
                      oscillation phases.

           \item[(b)] oscillations would not be observed if only the muon is
                      detected
        
            \item[(c)] neutrino oscillations can be observed even if the muon is
                      not detected.

            \end{itemize}
           Conclusion (b) is a correct consequence of the (incorrect) assumption
            that the muons recoiling against the different neutrino mass eigenstates
           have the same velocity. As both muons have the same mass they will have 
            equal proper time increments. So according to Eqn(2.12) the phase
           increments will also be equal and the interference term will vanish. 
           The conclusion (c) is in contradiction to the prediction of Eqn(3.22)
           of Ref.~\cite{SWS}, according to which, no neutrino oscillations can be observed
           when the decay position of muon is integrated over. The path amplitude 
           calculation of the present paper shows that conclusion (b) is no longer valid
           when the different possible times of propagation of the recoiling muons
           are taken into account.

         \par Observation of neutrino oscillations following pion decay,
          using a covariant formalism( Schwinger's parametric integral 
         representation of the space-time propagator) was considered in Ref.~\cite{Shtanov}.
         Exact energy-momentum conservation was imposed, but integration over the
         pion spatial position at amplitude level, as in Eqn(5.17), was done and no
         account was taken of the contribution of the pion space-time propagator to
         the oscillation phase. As the different eigenstates were assumed to be
          be produced or detected at identical space-time points equal propagation
         velocities were implicitly assumed, so that just as in Refs\cite{LDR,SWS,DMOS},
          where the same assumption was made, the standard neutrino oscillation 
          phase was obtained. In the conclusion of this paper the almost 
          classical nature of the space-time trajectories followed by the neutrinos
          was stressed, although this was not taken to its logical conclusion
          in the previous discussion, {\it e.g.} the kinematical inconsistency of the
          equal velocity hypothesis that requires the evidently impossible
          condition (5.18) to be satisfied. 
         \par In a recent paper~\cite{Ohlsson}, the standard neutrino oscillation formula
          with oscillation phase given by Eqn(1.1) was compared with a neutrino decoherence
          model. In order to take into account incertainties in the position of the source
          and the neutrino energy, an average was made over the quantity $L/4E_{\nu}$, assuming
          that it is distributed according to a Gaussian with mean value $\ell$ and width
          $\sigma$. The average was performed in an incoherent manner. Thus the calculation
          is closely analagous to those for the effects of target or detector length or
          of thermal motion of the neutrino source, presented in the Appendix A of the 
          present paper. Perhaps uniquely then, in the published literature, in
          Ref.~\cite{Ohlsson} the effects of source position and motion are taken into
          account correctly, according to Eqn(5.16) instead of Eqn(5.17). However, the
          source of the neutrino energy uncertainty is not specified. In as far as it
          is generated from source motion the calculation is, in principle, correct. 
          There is however also the (typically much smaller, see Table 1 above, for the 
          case of pion decay at rest) coherent contribution originating from the
          variation in the physical masses of the unstable particles produced in
          association with the neutrino, as discussed in detail above. It was concluded
          in Ref.~\cite{Ohlsson} that the Gaussian averaging procedure used gave
          equivalent results to the decoherence model for a suitable choice of parameters.
    
          \par It is clearly of great interest to apply the calculational method developed 
           in the present paper to the case of neutral kaon and b-meson oscillations.
           Indeed the use of the invariant path amplitude formalism has previously 
           been recommended~\cite{KaySto} for experiments involving correlated
           pairs of neutral kaons. Here, just a few remarks will be made on the main
           differences to be expected from the case of neutrino or muon oscillations.
           A more detailed treatment will be presented elsewhere\cite{JHF}.
           \par In the case of neutrino and muon oscillations, the interference effect
            is possible as the different neutrino eigenstates can be produced 
           at different times. This is because the decay lifetimes of all
           interesting sources (pions, muons, $\beta$-decaying nuclei) are much longer
           than the time difference beween the paths corresponding to the interfering
            amplitudes. To see
           if a similar situation holds in the case of $\rm{K}_S-\rm{K}_L$ oscillations, three
           specific examples will be considered with widely differing momenta of the
            neutral kaons:
            \begin{itemize}
            \item[(I)] $\phi \rightarrow \rm{K}_S \rm{K}_L$
            \item[(II)] $\pi^-p \rightarrow \Lambda \rm{K}^0$ at $\sqrt{s} = 2$ GeV 
            \item[(III)] $\pi^-p \rightarrow \Lambda \rm{K}^0$ at $\sqrt{s} = 10$ GeV. 
            \end{itemize}
             These correspond to neutral kaon momenta of 108 MeV, 750 MeV and 5 GeV
             respectively. In each case the time difference ($\Delta t_K$) of
             production of
             $\rm{K}_S$ and $\rm{K}_L$ mesons, in order that they arrive at the same time 
             at a point distant $c\gamma_K \tau_S$ (where $\gamma_K$ is the
             usual relativistic parameter) from the source is calculated. Exact
             relativistic kinematics is assumed and only leading terms in 
             the mass difference $\Delta m_K = m_S-m_L$ are retained. Taking
             the value of  $\Delta m_K$ and the various particle masses from
             Ref.~\cite{PDG} the following results are found for $\Delta t_K$  in the
             three cases: (I) 3.4$\times 10^{-23}$sec, (II) 8.3$\times 10^{-25}$sec  
             and (III) 6.4$\times 10^{-26}$sec. For comparison, for neutrino oscillations
             following pion decay at rest, with $\Delta m^2 = (1\rm{eV})^2$ and $L= 30$ m,
             Eqn(2.29) gives $\Delta t_{\nu} = 5.6\times 10^{-23}$sec. The result (I)
             may be compared with the mean life of the $\phi$ meson of 1.5$\times 10^{-22}$sec
             ~\cite{PDG}. Thus the $\phi$ lifetime is a factor of about five larger than
             $\Delta t_K$ indicating that $\rm{K}_S-\rm{K}_L$ interference should be possible by
              a similar mechanism to neutrino oscillations following pion decays,
              {\it i.e.} without invoking velocity smearing of the neutral kaon mass
              eigenstates. In cases (I) and (II) the interference effects observed will
              depend on the `characteristic time' of the (non resonant) strong 
              interaction process, a quantity that has hitherto not been susceptible
              to  experimental investigation~\footnote{A similar physical quantity
               has been considered in
                 Ref.~\cite{DeLeoRot}, where the possiblity of observable 
               modifications to the exponential decay law and the Breit-Wigner line
                shape distribution is suggested.}. If this time is much less than, or comparable
              to, $\Delta t_K$, essentially equal velocities (and therefore appreciable
               velocity smearing) of the eigenstates will be necessary for interference
              to occur. Since $\Delta m_K$ and $\Gamma_S$ are comparable in size, velocity
              smearing effects are expected to be, in any case, much larger than for neutrino
              oscillations following pion decay. These effects are readily calculated
              using the Gaussian approximation (2.22) of the present paper. The main
              contribution to the velocity smearing is due to the variation of the
              physical mass of the $\rm{K}_S$ rather than those of the $\rm{K}_L$ or
               $\Lambda$. 
              \par For the $\rm{B}_1-\rm{B}_2$ oscillation case, analagous to (I) above,
              $\Upsilon(4S) \rightarrow \rm{B}_1\rm{B}_2$ ($p_B= 335$ MeV) the value of
              $\Delta t_B$ is found to be $1.8\times 10^{-22}$sec, to be compared
              with $\tau(\Upsilon(4S)) = 4.7\times 10^{-23}$sec~\cite{PDG}, which is a factor
              3.8 smaller. Thus, velocity smearing effects are expected to play an important
              role in $\rm{B}_1-\rm{B}_2$ oscillations. This is possible, since the neutral b-meson
              decay width ($4.3\times10^{-10}$ MeV), and mass difference ($3.1\times10^{-10}$ MeV),
              have similar sizes. 
             
              \par In closing, it is interesting to mention two types of atomic
              physics experiments where interference effects similar to the
              conjectured (and perhaps observed~\cite{LNSD,KajTot,BahPinBas})
               neutrino oscillations have aleady
              been clearly seen.         
              \par The first is quantum beat spectroscopy~\cite{BerSub}. This type
               of experiment, which has previously been discussed in connection
               with neutrino oscillations~\cite{SW}, corresponds closely to the
               gedanken experiment used by Heisenberg~\cite{Heis} to exemplify the
               the fundamental law of quantum mechanics Eqn(1.2). The atoms of an
               atomic beam are excited by passage through a thin foil or a laser beam.
               The quantum phase of an atom with excitation energy $E^*$ evolves
               with time according to: $\exp(-iE^*\Delta t)$ (see the discussion after
               Eqn(3.3) above). If decay photons from two nearby states with 
               excitation energies $E^*_{\alpha}$ and $E^*_{\beta}$ are detected after 
               a time interval $\Delta t$ ( for example by placing a photon detector
               beside the beam at a variable distance $d$ from the excitation foil) a 
               cosine interference term with phase: 
               \begin{equation}
               \phi_{beat} = \frac{(E^*_{\alpha}-E^*_{\beta})d}{\overline{v}_{atom}}
               \end{equation} 
                where $\overline{v}_{atom}$ is the average velocity of the atoms in 
                the beam, is observed~\cite{BerSub}. An atom in the beam, before
                excitation, corresponds to the neutrino source pion. The excitation
                process corresponds to the decay of the pion. The propagation of the
                two different excited states, {\it alternative} histories of the
                initial atom, correspond to the {\it alternative} propagation of
                the two neutrino mass eigenstates. Finally the dexcitation of 
                the atoms and the detection of a {\it single} photon corresponds 
                to the neutrino detection process. The particular importance
                of this experiment for the path amplitude calculations
                presented in the present paper, is that it demonstrates, experimentally,
                the important contribution to the interference phase of the space-time
                propagators of excited atoms, in direct analogy to the similar
                contributions of unstable pions, muons and nuclei discussed above.

                \par An even closer analogy to neutrino osillations following
                 pion decay is provided by the recently observed process of
                 photodetachement of an electron by laser excitation:
                 the `Photodetachment Microscope'~\cite{BDD}. A laser photon ejects
                 the electron from, for example, an $^{16}\rm{O}^-$ ion in a beam. The 
                 photodetached electron is emitted in an S-wave (isotropically)
                 and with a fixed initial energy. It then moves in a constant,
                 vertical, electric field that is perpendicular to the direction
                 of the ion beam and almost parallel to the laser beam. An upward
                moving electron that is decelerated by the field eventually
                undergoes `reflection' before being accelerated towards a planar
                position-sensitive electron detector situated below the beam
                and perpendicular
                to the electric field direction ( see Fig.1 of Ref.~\cite{BDD}). In these
                 circumstances, it can be shown~\cite{BBGKM} that, just two classical 
                 electron trajectories link the production point to any point in the
                 kinematically allowed region of the detection plane. Typical parameters
                 for $^{16}\rm{O}^-$ are~\cite{BBD}: initial electron kinetic energy, 102 keV;
                  detector distance, 51.4 cm; average time-of-flight, 117 ns; difference
                  in emission times to arrive in spatial-temporal coincidence at the
                  detector plane, 160 ps. An interference pattern is generated by the
                  phase difference between the amplitudes corresponding to the two
                  allowed trajectories. The phase difference, derived by performing
                  the Feynman path integral of the classical action along the 
                  classical trajectories~\cite{BBD}, gives a very good description
                  of the observed interference pattern. The extremely close analogy
                 between this experiment and the neutrino oscillation experiments
                 described in Sections 2 and 3 above is evident. Notice that the
                 neutrinos, like the electrons in the photodetachement
                 experiment, must be emitted at different times, in the 
                  alternative paths, for interference to be possible. This is 
                 the crucial point that was not understood in previous
                 treatments of the quantum mechanics of neutrino oscillations.
                  \par Actually, Ref.~\cite{BBD} contains, in Section IV, a path
                 amplitude calculation for electrons in free space that is
                 geometrically identical to the discussion of pion decays in
                 flight presented Section 4 above (compare Fig.3 of the present
                 paper with Fig.3 of  Ref.~\cite{BBD}) The conclusion of
                 Ref.~\cite{BBD} is that, in this case, no interference effects
                 are possible for electrons that are mononergetic in the
                 source rest frame. As is shown in Section 4 above, if these
                 electrons are replaced either by neutrinos of different
                 masses from pion decay, or muons recoiling against such neutrinos,
                 observable interference effects are indeed to be expected. 
\newpage

{\bf Appendix A}
\par Random thermal motion of the decaying pion in the target has two distinct
 physical effects on the phase of neutrino oscillations,
~~~~~~~~~~~~~~~~~~~~~~~~~~~~~~~~~ \[ \phi_{12}^{\nu,\pi}(0) = 
-\frac{\Delta m^2 L}{P_0}+\frac{m_{\pi}\Delta m^2 L}{2P_0^2}
~~~~~~~~~~~~~~~~~~~~~~~~~~~~~~(A1) \]
(where the first and second terms in Eqn(A1) give the contributions of
 the neutrino and pion paths respectively):
\begin{itemize}
\item[(1)] The observed neutrino momentum, $P_{\nu}$, is no longer equal to
 $P_0$, due to the boost from the pion rest frame to the laboratory system.
 (Doppler effect or Lorentz boost)
\item[(2)] The time increment of the pion path $t_D-t_0$ (see Eqn(2.16)) no longer
 corresponds to the pion proper time. (Relativistic time dilatation)
\end{itemize}
Taking into account (1) and (2) gives, for the neutrino oscillation phase:
~~~~~~~~~~~~~~~~~~~~~~~~~~~~~~~~~~ \[ \phi_{12}^{\nu,\pi}(corr) = 
-\frac{\Delta m^2 L}{P_{\nu}}+\frac{m_{\pi}\Delta m^2 L}{2 \gamma_{\pi} P_{\nu}^2}
~~~~~~~~~~~~~~~~~~~~~~~~~~~~(A2) \]
where:
\[ P_{\nu} = \gamma_{\pi} P_0(1+v_{\pi}\cos \theta_{\nu}^{\ast})~~{\rm and}
~~\gamma_{\pi}=\frac{E_{\pi}}{m_{\pi}} \]
Here, $\theta_{\nu}^{\ast}$ is the angle between the neutrino momentum vector
 and the pion flight direction in the pion rest frame.
 Developing $\gamma_{\pi}$ and $v_{\pi}$ in terms of the small quantity $p_{\pi}/m_{\pi}$,
 Eqn(A2) may be written as:
 \[ \phi_{12}^{\nu,\pi}(corr) =  \phi_{12}^{\nu,\pi}(0)+
\frac{p_{\pi}}{m_{\pi}}\frac{\Delta m^2 L}{P_0}\left[1-\frac{m_{\pi}}{P_0}\right]
\cos\theta_{\nu}^{\ast}
+ \left(\frac{p_{\pi}}{m_{\pi}}\right)^2 \frac{\Delta m^2 L}{2 P_0}\left[1-\frac{3 m_{\pi}}{2 P_0}\right]   
~~~~~~~(A3) \]
Performing now the average of the interference term over the isotropic distribution
 in $ \cos\theta_{\nu}^{\ast}$:
\begin{eqnarray}
 \langle \cos \phi_{12}^{\nu,\pi}(corr) \rangle_{\theta_{\nu}^{\ast}}
  & = & \frac{1}{4}{\it Re}\int_{-1}^{1} \exp\left[i\phi_{12}^{\nu,\pi}(corr)\right]
 d \cos \theta_{\nu}^{\ast}  \nonumber \\
  & = & \frac{1}{2} {\it Re} \exp \left\{i \phi_{12}^{\nu,\pi}(0)+
\left(\frac{p_{\pi}}{m_{\pi}}\right)^2\left(i\frac{\Delta m^2 L}{2 P_0}
\left[1-\frac{3 m_{\pi}}{2 P_0}\right] \right. \right.  \nonumber \\ 
 &  & \left. \left.
~~~~~~~~~~-\frac{1}{6}\left(\frac{\Delta m^2 L}{P_0}
\left[1-\frac{m_{\pi}}{P_0}\right] \right)^2\right) 
\right\}~~~~~~~~~~~~~~~~~~(A4) \nonumber
\end{eqnarray}
In deriving Eqn(A4) the following approximate formula is used:
~~~~~~~~~~~~~~~~~~~~\[ \frac{1}{2}\int_{-1}^{1}e^{i\alpha c}dc =
 \frac{1}{2i\alpha}\left[e^{i\alpha}-e^{-i\alpha}\right]
= \frac{\sin \alpha}{\alpha} \simeq 1-\frac{\alpha^2}{6}~~~~~~~~~~~~~~~~~~~~~~~~~(A5) \]
where 
\[ \alpha \equiv \frac{p_{\pi}}{m_{\pi}}\frac{\Delta m^2 L}{P_0}
\left[1-\frac{m_{\pi}}{P_0}\right] \ll 1 \]
The average over the Maxwell-Boltzmann distribution (2.31) is readily performed by
 `completing the square' in the exponential, with the result:
\begin{eqnarray}
 \langle \cos \phi_{12}^{\nu,\pi}(corr) \rangle_{\theta_{\nu}^{\ast},p_{\pi}}
 & = &\frac{1}{2}{\it Re}\exp\left\{-\left(\frac{\overline{p}_{\pi}\Delta m^2 L}{2 m_{\pi} P_0}
\left[1-\frac{m_{\pi}}{P_0}\right]\right)^2 \right. \nonumber \\
 &   &  + \left. i\left[\phi_{12}^{\nu,\pi}(0)+
\frac{3}{4}\left(\frac{\overline{p}_{\pi}}{m_{\pi}}\right)^2\left(\frac{\Delta m^2 L}{P_0}
\left[\frac{3 m_{\pi}}{2 P_0}-1\right]\right)\right]\right\} \nonumber \\
 & \equiv &  F^{\nu}(Dop)\cos [\phi_{12}^{\nu,\pi}(0)+\phi^{\nu}(Dop)]
\nonumber~~~~~~~~~~~~~~~~~~~~~~~~~~~~~~~~(A6)
\end{eqnarray}
leading to Eqns(2.32) and (2.33) for the Doppler damping factor $F^{\nu}(Dop)$ and phase 
 shift $\phi^{\nu}(Dop)$, respectively.
\par The correction for the effect of thermal motion in the case of muon oscillations
 $\phi^{\nu}(Dop)$, respectively. is performed in a similar way. The oscillation phase:
~~~~~~~~~~~~~~~~~~~~~~~~~~~~~~~~~ \[ \phi_{12}^{\mu,\pi}(0) = 
-\frac{m_{\mu}^2 E_0^{\mu}\Delta m^2 L}{2 m_{\pi} P_0^3}+\frac{m_{\mu}\Delta m^2 L}{2P_0^3}
~~~~~~~~~~~~~~~~~~~~~~~~(A7) \]
 is modified by the Lorentz boost of the muon momentum and energy, and the relativistic 
 time dilatation of the phase increment of the pion path, to:
~~~~~~~~~~~~~~~~~~~~~~~~~~~~~~~~~\[ \phi_{12}^{\mu,\pi}(corr) = 
-\frac{m_{\mu}^2 E_{\mu}\Delta m^2 L}{2 m_{\pi} P_{\mu}^3}+\frac{m_{\mu}\Delta m^2 L}
{2 \gamma_{\pi}P_{\mu}^3}~~~~~~~~~~~~~~~~~~~~(A8) \]
where
\[ E_{\mu} = \gamma_{\pi}E_0^{\mu}(1+v_{\pi}v_0^{\mu} \cos \theta_{\mu}^{\ast}) \]
and $v_0^{\mu}$ is given by Eqn(2.37). Developing, as above, in terms of $p_{\pi}/m_{\pi}$, 
gives:
 \[ \phi_{12}^{\mu,\pi}(corr) =  \phi_{12}^{\mu,\pi}(0)+
\frac{p_{\pi}}{m_{\pi}}\frac{v_0^{\mu} m_{\mu}^2 \Delta m^2 L}{P_0^3}
\left[\frac{E_0^{\mu}}{m_{\pi}}-\frac{3}{2}\right]
\cos\theta_{\mu}^{\ast}
+ \left(\frac{p_{\pi}}{m_{\pi}}\right)^2 \frac{m_{\mu}^2 \Delta m^2 L}{P_0^3}
\left[\frac{E_0^{\mu}}{2 m_{\pi}}-1\right]   
(A9) \] 
Performing the averages over $\theta_{\mu}^{\ast}$ and $p_{\pi}$ then leads to 
Eqns(2.49) and (2.50) for the damping factor $F^{\mu}(Dop)$ and phase 
 shift $\phi^{\mu}(Dop)$, respectively.
\par The effect of the finite longitudinal dimensions of the target or detector is calculated
 by an appropriate weighting of the interference term according to the value of the
 distance $X=x_f-x_i$ between the decay and detection points (see Fig.1). 
 Writing the interference phase as $\phi_{12}=\beta X$, and assuming a uniform 
 distribution of decay points within the target of thickness $\ell_T$:
\begin{eqnarray}
~~~~~~~~~~~~~~~~~~~~~~~~~~~~~~\langle \cos \phi_{12}
 \rangle & = & \frac{1}{\ell_T}\int_{L-\frac{\ell_T}{2}}
^{L+\frac{\ell_T}{2}}\cos \beta X dX \nonumber \\
 & = & \frac{2}{\beta \ell_T}\sin\frac{\beta \ell_T}{2} \cos \beta L  \nonumber \\
 & \equiv & F_{Targ} \cos \beta L
\nonumber~~~~~~~~~~~~~~~~~~~~~~~~~~~~~~~~(A10)
\end{eqnarray}
 Substituting the value of $\beta$ appropriate to neutrino oscillations yields
 Eqn(2.34). Since the value of $\beta$ is the same for neutrino and muon oscillations,
 the same formula is also valid in the latter case. The same correction factor, with
 the replacement $\ell_T \rightarrow \ell_D$ describes the effect of a finite detection
 region of length $\ell_D$:
\[ L-\frac{\ell_D}{2}+x_i < x_f < L+\frac{\ell_D}{2}+x_i \]      
\newpage

{\bf Appendix B}
\par The first step in the derivation of Eqn(4.17) relating $\Delta v(\mu)$ to $\Delta m^2$
is to calculate the angle $\delta^{\ast}$, in the centre-of-mass (CM) system of the decaying
 pion, corresponding to $\delta$ in the laboratory (LAB) system (see Fig.3). It is assumed,
 throughout,
 that the pion and muon are ultra-relativistic in the latter system, so that:
 $v_{\pi},v_i(\mu) \simeq 1$. The Lorentz transformation relating the CM and LAB systems
 gives the relation:
\[ \sin \theta_i = \frac{v_i^{\ast}(\mu) \sin \theta_i^{\ast}}
   {\gamma_{\pi}(1+v_i^{\ast}(\mu)\cos \theta_i^{\ast})}~~~~~i=1,2
~~~~~~~~~~~~~~~~~~~~(B1) \]
 The starred quantities refer to the pion CM system. Making the substitutions:
 $\theta_2 = \theta_1+\delta$, $\theta_2^{\ast} = \theta_1^{\ast}+\delta^{\ast}$, 
 Eqns(B1) may be solved to obtain, up to first order in $\delta$, $\delta^{\ast}$
 and $\Delta m^2$:
\[ \Delta v^{\ast}(\mu) = v_2^{\ast}(\mu)-v_1^{\ast}(\mu) = 
 \frac{\gamma_{\pi}(1+ v_0^{\ast}(\mu) \cos \theta^{\ast}_1)^2\delta-
v_0^{\ast}(\mu)(\cos \theta^{\ast}_1+v_0^{\ast}(\mu))\delta^{\ast}}
{\sin \theta^{\ast}_1}~~~~~~~(B2) \]
where, (c.f. Eqn(2.37):
\[ v_0^{\ast}(\mu)=\frac{m_{\pi}^2-m_{\mu}^2}{m_{\pi}^2+m_{\mu}^2}
~~~~~~~~~~~~~~~~~~~~~~~~~~~~~~~~~~~~~~~(B3) \]
Using Eqn(2.36) $\Delta v^{\ast}(\mu)$ may be expressed in terms of the neutrino mass
 difference:
 \[ \Delta v^{\ast}(\mu) = \frac{4 m_{\mu}^2 m_{\pi}^2 \Delta m^2}{(m_{\pi}^2-m_{\mu}^2)
(m_{\pi}^2+m_{\mu}^2)^2}
~~~~~~~~~~~~~~~(B4)  \]
 Eliminating now $\Delta v^{\ast}(\mu)$ between (B2) and (B4) gives a relation between 
 $\delta$, $\delta^{\ast}$ and $\Delta m^2$:  
\[\delta^{\ast} = \frac{\gamma_{\pi}(1+ v_0^{\ast}(\mu) \cos \theta^{\ast}_1)^2 \delta}
 {v_0^{\ast}(\mu)(\cos \theta^{\ast}_1+v_0^{\ast}(\mu))}-
\frac{4 m_{\mu}^2 m_{\pi}^2 \Delta m^2 \sin \theta^{\ast}_1 }{(m_{\pi}^2-m_{\mu}^2)^2
(m_{\pi}^2+m_{\mu}^2) (\cos \theta^{\ast}_1+v_0^{\ast}(\mu))}~~~~~~~~~~(B5) \]
 In the LAB system, and in the UR limit, the difference of the velocities of the
 muons recoiling against the two neutrino mass eigenstates is:
 \[ \Delta v(\mu) = v_2(\mu)- v_1(\mu) = \frac{P_2(\mu)}{E_2(\mu)}-\frac{P_1(\mu)}{E_1(\mu)}
\simeq \frac{m_{\mu}^2}{E_{\mu}^3}[E_2(\mu)-E_1(\mu)]~~~~~~~~~~~~~~~~~(B6) \]
where $E_{\mu}$ is the muon energy in the LAB system for vanishing neutrino masses.
Making the Lorentz transformation of the muon energy from the pion CM to the LAB
 frames, and using Eqns(2.4) and (2.36) to retain only terms linear in $\Delta m^2$
 and $\delta^{\ast}$, enables Eqn(B6) to be re-written as:
\[\Delta v(\mu) = \frac{E_{\pi}}{2 E_{\mu}^3}\left(\frac{m_{\pi}}{m_{\mu}}\right)^2 
 \left[\frac{ \Delta m^2  (\cos \theta^{\ast}_1+v_0^{\ast}(\mu))}
 {v_0^{\ast}(\mu)}- \delta^{\ast}(m_{\pi}^2-m_{\mu}^2)\sin \theta^{\ast}_1\right]
~~~~~~~~~~~~~~~~~~~~~~~(B7) \]
where $E_{\pi}$ is the energy of the pion beam. By combining the geometrical constraint
 equation for the muon velocities, (4.16) with (B5) and (B7) the angles $\delta$ and
 $\delta^{\ast}$ may be eliminated to yield the equation for LAB frame velocity
 difference:
\[\Delta v(\mu) = \frac{E_{\pi} \Delta m^2}{2 m_{\pi}^2 (m_{\pi}^2-m_{\mu}^2)} \frac{A}{B}
~~~~~~~~~~~~~~~~~~~~~~~~~~~(B8) \]
 where 
\[ A = (v_1(\mu)-v_{\pi} \cos \theta_1)
 \left\{(\cos \theta^{\ast}_1+v_0^{\ast}(\mu))^2+\frac{4 m_{\mu}^2 m_{\pi}^2
  \sin^2 \theta^{\ast}_1 }{(m_{\pi}^2+m_{\mu}^2)^2} \right\}~~~~~~~~(B9) \]
 \begin{eqnarray}
 B & = & \frac{E_{\pi}(m_{\pi}^4-m_{\mu}^4)(1+ v_0^{\ast}(\mu) \cos \theta^{\ast}_1)}
 {8 m_{\pi}^4  m_{\mu}^2}  \nonumber \\
  &  & \times \left\{\frac{E_{\pi}^2 (m_{\pi}^2+m_{\mu}^2)(1+ v_0^{\ast}(\mu) \cos \theta^{\ast}_1)^2
 (\cos \theta^{\ast}_1+v_0^{\ast}(\mu))(v_1(\mu)-v_{\pi} \cos \theta_1)}
 {m_{\pi}^2 (m_{\pi}^2-m_{\mu}^2)} \right. \nonumber \\
  &  & + \left. \frac{4 m_{\mu}^2 m_{\pi}^2 
  \sin^2 \theta^{\ast}_1 }{(m_{\pi}^2+m_{\mu}^2)^2} \right\}
~~~~~~~~~~~~~~~~~~~~~~~~~~~~~~~~~~~~~~~~~~~~~~~~~~~~~~~~~~~~~~~(B10) \nonumber 
\end{eqnarray}
 \par To simplify (B8), the quantity $(v_1(\mu)-v_{\pi} \cos \theta_1)$ is now expressed in terms of
 kinematic quantities in the pion CM system. Within the UR approximation used,
 \[ \theta_1, m_{\pi}/E_{\pi}, m_{\mu}/E_{\mu} \ll 1   \]
 so that 
\[ v_1(\mu)-v_{\pi} \cos \theta_1 = \frac{1}{2}\left(\frac{m_{\pi}^2}{E_{\pi}^2}-\frac{m_{\mu}^2}{E_{\mu}^2}
 + \theta_1^2 \right) + O( \left(\frac{m_{\pi}}{E_{\pi}}\right)^4, \left(\frac{m_{\mu}}{E_{\mu}}\right)^4,
  \theta_1^4)~~~~~~~~~~~~~(B11) \]
Writing Eqn(B1) to first order in $\theta_1$, and neglecting terms of $O(\theta_1 m_i^2)$:
\[~~~~~~~~~~ \theta_1 = \frac{ m_{\pi} v_0^{\ast} \sin \theta^{\ast}_1 }{E_{\pi}
(1+ v_0^{\ast}(\mu) \cos \theta^{\ast}_1)}
~~~~~~~~~~~~~~~~~~~~~~~~~~~~~~~~~~~~~~~~~~~~~(B12) \]
 Using Eqn(B12), and expressing $E_{\mu}$ in terms of pion CM quantities, Eqn(B11) may 
 be written as:
\[  v_1(\mu)-v_{\pi} \cos \theta_1 = \frac{m_{\pi}^2(m_{\pi}^2-m_{\mu}^2) 
(\cos \theta^{\ast}_1+v_0^{\ast}(\mu))}
{E_{\pi}^2 (m_{\pi}^2+m_{\mu}^2)(1+ v_0^{\ast}(\mu) \cos \theta^{\ast}_1)^2}
~~~~~~~~~~~~~~~~~(B13) \]
Expressing the RHS of (B13) in terms of $E_{\pi}$ and $E_{\mu}$, using the relation:
\[ \cos \theta^{\ast}_1 = \frac{m_{\pi}^2(2E_{\mu}-E_{\pi})-m_{\mu}^2 E_{\pi}}
 { E_{\pi}(m_{\pi}^2-m_{\mu}^2)}~~~~~~~~~~~~~~~~~~~~~~~~~~~~~~~(B14) \]
gives Eqn(4.20) of the text.
\par On substituting (B13) into the RHS of (B10), it can be seen that the factor in the
 large curly brackets is the same in (B9) and (B10), and so cancels in the ratio $A/B$ 
 in Eqn(B8). It follows that:
 \[ \Delta v(\mu) = \frac{m_{\mu}^2 \Delta m^2}{E_{\mu}^2 (m_{\pi}^2-m_{\mu}^2)}
\left(\frac{\cos \theta^{\ast}_1+v_0^{\ast}(\mu)}{1+ v_0^{\ast}(\mu) \cos \theta^{\ast}_1}
\right)~~~~~~~~~~~~~~~~~~
~~~~~~~~~~(B15)  \]
 Finally, using (B3) and (B14) to express the factor in large brackets in Eqn(B15)
 in terms of $E_{\mu}$ and $E_{\pi}$, Eqn(4.17) of the text is obtained.

\vspace*{4cm}
\begin{figure}[htbp]
\begin{center}
\hspace*{-0.5cm}\mbox{
\epsfysize15.0cm\epsffile{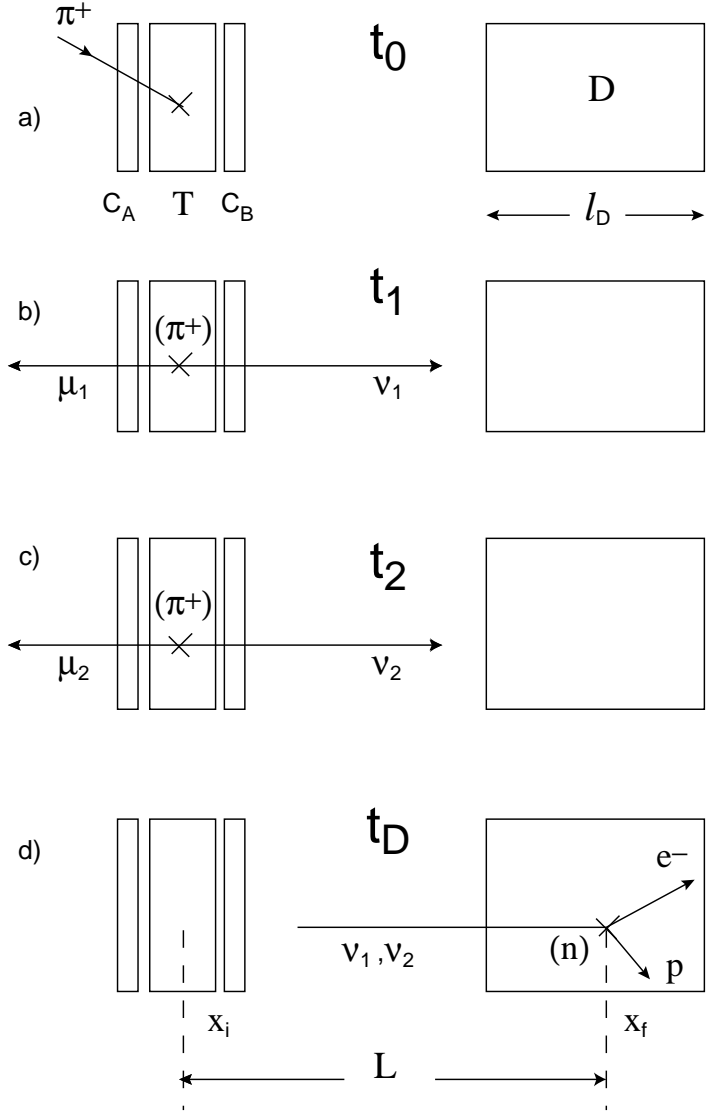}}
\caption{The space-time description of  $\nu_{\mu} \rightarrow \nu_e$ oscillations
  following $\pi^+$ decay at rest, in Feynman's formulation of quantum 
   mechanics. In a) a $\pi^+$ comes to rest in the stopping target T at time $t_0$.
   The pion, at rest at time $t_0$, constitutes the initial state for the path
   amplitudes. In b) and c) are shown two alternative classical histories for
   the $\pi^+$; in b),[~c)]  the pion decays into the mass eigenstate $|\nu_1>$,
[ $|\nu_2>$ ] at times $t_1$, [ $t_2$ ]. If $m_1~>~m_2$, and for suitable values
  of $t_1$ and $t_2$  ($t_2 > t_1$), the two classical histories may correspond
   to a common final state, shown in d) where the neutrino interaction
   $\nu_e n \rightarrow e^- p$ occurs at time $t_D$. As the initial and
   final states of the two classical histories are the same, the corresponding
   path amplitudes must be added coherently, as in Eqn(1.2), to calculate the
   probability of the whole process.}
\label{fig-fig1}
\end{center}
 \end{figure}
  
\vspace*{4cm}
\begin{figure}[htbp]
\begin{center}
\hspace*{-0.5cm}\mbox{
\epsfysize15.0cm\epsffile{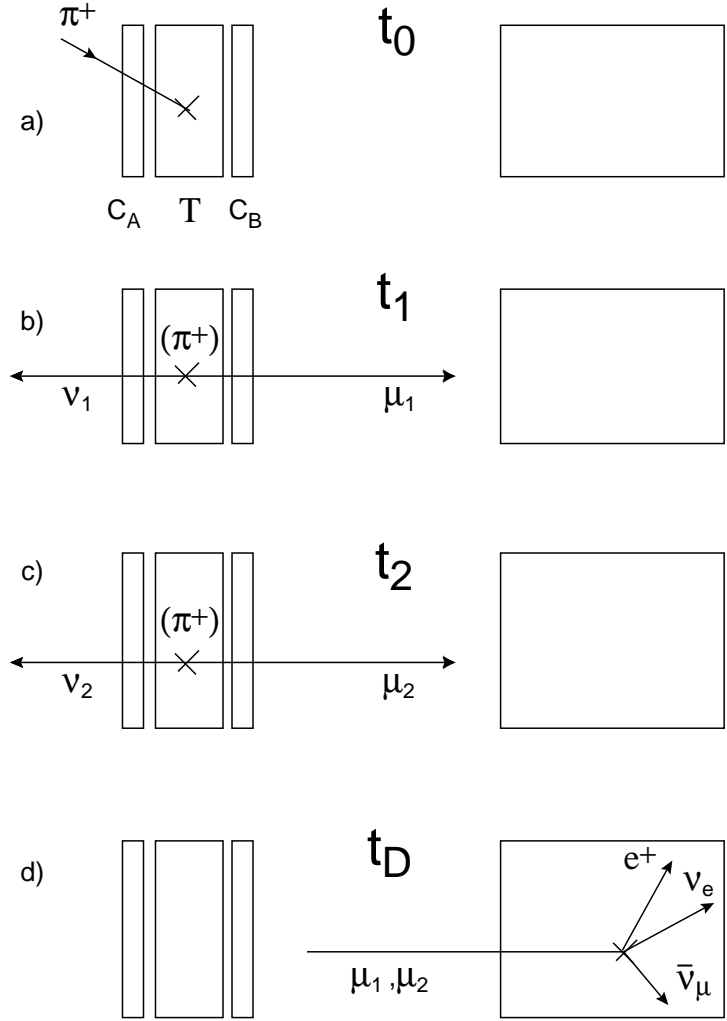}}   
\caption{The space-time description of `muon oscillations' induced by
   neutrino mass differences,
  following $\pi^+$ decay at rest, in Feynman's formulation of quantum 
   mechanics. As in Fig.1, b) and c) show alternative classical histories
   of the stopped $\pi^+$. If $m_1 > m_2$ the velocity of $\mu_1$ is
    less than that of $\mu_2$, and provided that $t_2 > t_1$, the muons
    may arrive at the same spatial point at the same time $t_D$ in both classical
    histories. If the muons are detected at this space-time point in
    any way (not necessarily by the observation of muon decay as shown
    in c)) interference between the correponding path amplitudes occurs,
    according to Eqn(1.2), just as in the case of neutrino detection.}
\label{fig-fig2}
\end{center}
 \end{figure}

\vspace*{4cm}
\begin{figure}[htbp]
\begin{center}
\hspace*{-0.5cm}\mbox{
\epsfysize15.0cm\epsffile{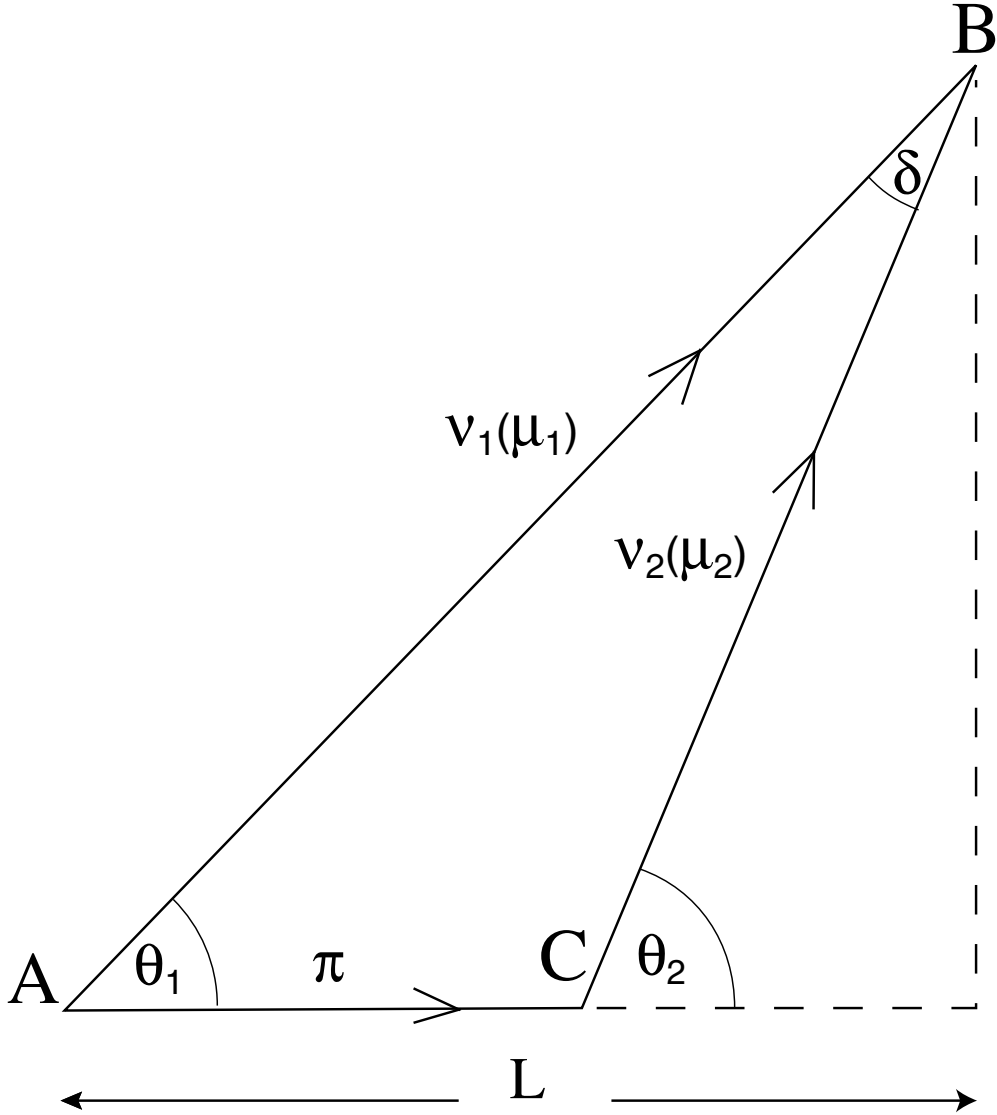}}  
\caption{ Two dimensional spatial geometry for the observation
      of neutrino or muon oscillations following pion decay in flight.
      Four possible classical histories of a pion, originally at the
      point A, are shown. In the first two, the pion decays either into 
     the mass eigenstate  $|\nu_1>$, at point A or into $|\nu_2>$ at 
     point C. If $m_1 > m_2$, and for suitable values of the angles
     $\theta_1$ and $\theta_2$, the neutrinos may arrive at the point
     B at the same time. If a neutrino detection event, such as
     $\nu_e n \rightarrow e^- p$, then occurs at B at this time,
     the amplitides
     corresponding to the paths AB and ACB will be indistinguishable
     so that they must be superposed, as in Eqn(1.2), to calculate
     the probability of the overall decay-propagation-detection 
     process. The third and fourth classical histories are 
     similar, except that the neutrino mass eigenstates
     are replaced by the corresponding recoil muons. The muons
     in the different histories may arrive at point B, at the same time,
     leading to interference and `muon oscillations' if
     they are detected there.}
\label{fig-fig3}
\end{center}
 \end{figure}
 
\end{document}